\documentclass{article}
\usepackage[english]{babel}
\usepackage{multicol}
\usepackage[math]{blindtext}
\usepackage{url}
\usepackage{amssymb,amsmath,subcaption}
\usepackage{amsmath}
\usepackage{amssymb}
\usepackage{dsfont}
\usepackage{psfrag}
\usepackage{hyperref}
\usepackage{multirow}
\usepackage{indentfirst}       
\usepackage{graphicx} 
\usepackage{url}              
\usepackage{textcomp}
\usepackage{xspace}
\usepackage{cite}
\usepackage{filecontents}
\usepackage{algorithm}
\usepackage{algpseudocode}
\usepackage{color}
\usepackage{footnote}
\usepackage{float}
\usepackage{booktabs}
\newcommand\blfootnote[1]{%
  \begingroup
  \renewcommand\thefootnote{}\footnote{#1}%
  \addtocounter{footnote}{-1}%
  \endgroup
}

\begin{document}

{\centering

{\bfseries\Large Distributed Multi-Speaker Voice Activity Detection for Wireless Acoustic Sensor Networks \bigskip}

Mohamad Hasan Bahari\textsuperscript{1}\blfootnote{The work of M.H. Bahari, J. Plata-chaves and M. Muma has been supported by the project HANDiCAMS which acknowledges the financial support of the Future and Emerging Technologies (FET) programme within the Seventh Framework Programme for Research of the European Commission, under FET-Open grant number: 323944.}, L. Khadidja Hamaidi\textsuperscript{2}\blfootnote{The work of L. K. Hamaidi is supported by the 'Excellence Initiative' of the German Federal and State Governments and the Graduate School of Computational Engineering at Technische Universit\"at Darmstadt.}
, Michael Muma\textsuperscript{2}, Jorge Plata-Chaves\textsuperscript{1}, Marc Moonen\textsuperscript{1}\blfootnote{The work of A. Bertrand and M. Moonen is supported by Research Project FWO nr. G.0931.14 'Design of distributed signal processing algorithms and scalable hardware platforms for energy-vs-performance adaptive wireless acoustic sensor networks'.}
, Abdelhak M. Zoubir\textsuperscript{2}, and Alexander Bertrand\textsuperscript{1}\\
  {\itshape
\textsuperscript{1} STADIUS Center for Dynamical Systems, Signal Processing and Data Analytics, Department of Electrical Engineering (ESAT), KU Leuven\\
  \textsuperscript{2}	Technische Universit\"at Darmstadt, Signal Processing Group, Darmstadt, Germany \\ 
  }
  \vspace{0.5cm}
\centerline{March 14, 2017}}

\begin{abstract}
A distributed multi-speaker voice activity detection (DM-VAD) method for wireless acoustic sensor networks (WASNs) is proposed. DM-VAD is required in many signal processing applications, e.g. distributed speech enhancement based on multi-channel Wiener filtering, but is non-existent up to date. The proposed method neither requires a fusion center nor prior knowledge about the node positions, microphone array orientations or the number of observed sources. It consists of two steps: (i) distributed source-specific energy signal unmixing (ii) energy signal based voice activity detection. Existing computationally efficient methods to extract source-specific energy signals from the mixed observations, e.g., multiplicative non-negative independent component analysis (MNICA) quickly loose performance with an increasing number of sources, and require a fusion center. To overcome these limitations, we introduce a distributed energy signal unmixing method based on a source-specific node clustering method to locate the 
nodes around each source. To determine the 
number of sources that are observed in the WASN,  a source enumeration method that uses a Lasso penalized Poisson generalized linear model is developed. Each identified cluster estimates the energy signal of a single (dominant) source by applying a two-component MNICA. The VAD problem is transformed into a clustering task, by extracting features from the energy signals and applying K-means type clustering algorithms. All steps of the proposed method are evaluated using numerical experiments.  A VAD accuracy of $> 85 \%$ is achieved for a challenging scenario where 20 nodes observe 7 sources in a simulated reverberant rectangular room.
\end{abstract}
%


\section{Introduction}
\begin{figure}[h]
	\centering
	\includegraphics[width=130mm]{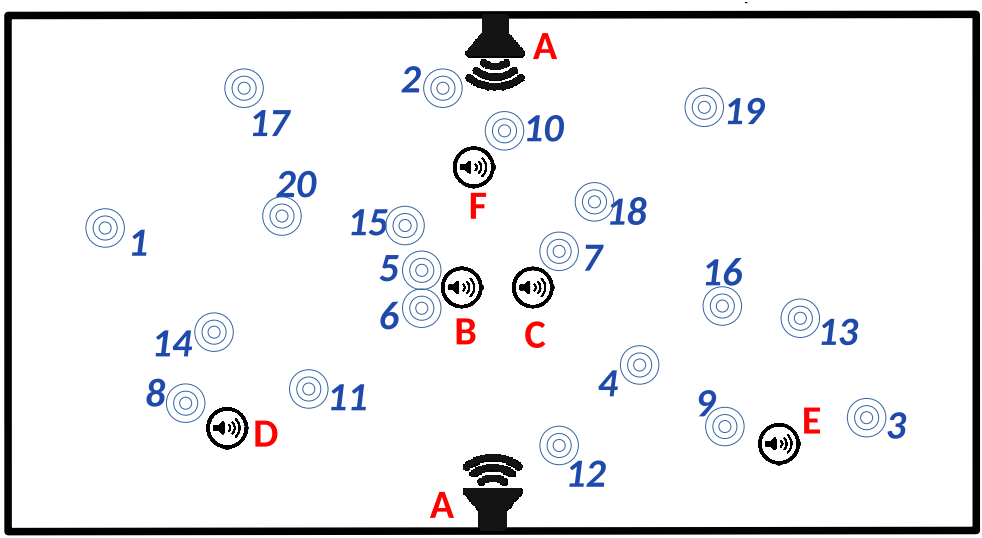}
	\caption{An example of a WASN observing seven speech sources (red) in a $20 \times 10$m room with reverberation time T$60=0.3$s. The microphone signals of the nodes (blue) are sampled at $16$kHz. Source A models a public address system playing the same announcement from two different loudspeakers. Sources B, C, D, E and F are five different speech sources.}    
	\label{fig:scenario}
\end{figure}

Wireless acoustic sensor networks (WASN) are composed of a multitude of spatially distributed nodes (see Fig. \ref{fig:scenario}) with wireless communication and computation capabilities, each containing a single microphone or a microphone array. By leveraging upon spatial diversity, WASNs allow for improved speech enhancement compared to single-node methods. Improving the quality and/or intelligibility of a speech signal corrupted by noise has a wide variety of applications, e.g. in hands-free telephony, teleconferencing, automatic speech recognition and hearing aids~\cite{bertrand2011distributed,lanbo2008prospects,ismail2012acoustic, chong2003sensor,bertrand2011applications,Bertrand2009}. WASNs may operate in a centralized configuration, where the nodes transmit their observations to a fusion center that performs all processing. Alternatively, decentralized WASNs do not require a fusion center as they distribute the computations among the nodes. In this way, the communication cost is reduced by 
communicating 
only within a neighborhood. Furthermore, robustness against a single point of failure (fusion center), as well as scalability to larger networks, are improved at the cost of an increased computational demand at each node~\cite{DANSE1,DANSE2,chouvardas2015distributed}. 

Recently, several distributed speech enhancement algorithms have been developed that are based on multi-channel Wiener filtering (MWF), such as the distributed adaptive node-specific signal estimation (DANSE) algorithm~\cite{DANSE1,DANSE2,bertrand2011distributed,hassani2015low}. These methods require distributed multi-speaker voice activity detection (DM-VAD) to estimate the speech and noise covariances. While single-speaker single-node VAD is a well researched problem \cite{VAD2,VAD3,VAD4,VAD5,VAD9,VAD11,VAD10,VAD6,VAD12,VAD13,verteletskaya2010voice}, to the best of our knowledge, no DM-VAD method is available in the literature. Even for centralized WASNs, the literature is sparse \cite{multivad3,multivad2,multivad1}.
Therefore, in this paper, a DM-VAD for WASNs is proposed. It consists of two steps: (i) distributed source-specific energy signal unmixing (ii) energy signal based voice activity detection. 

\begin{figure*}[t]
\begin{minipage}{0.33\linewidth}
  \centering
 \includegraphics[width=46 mm, height=35mm]{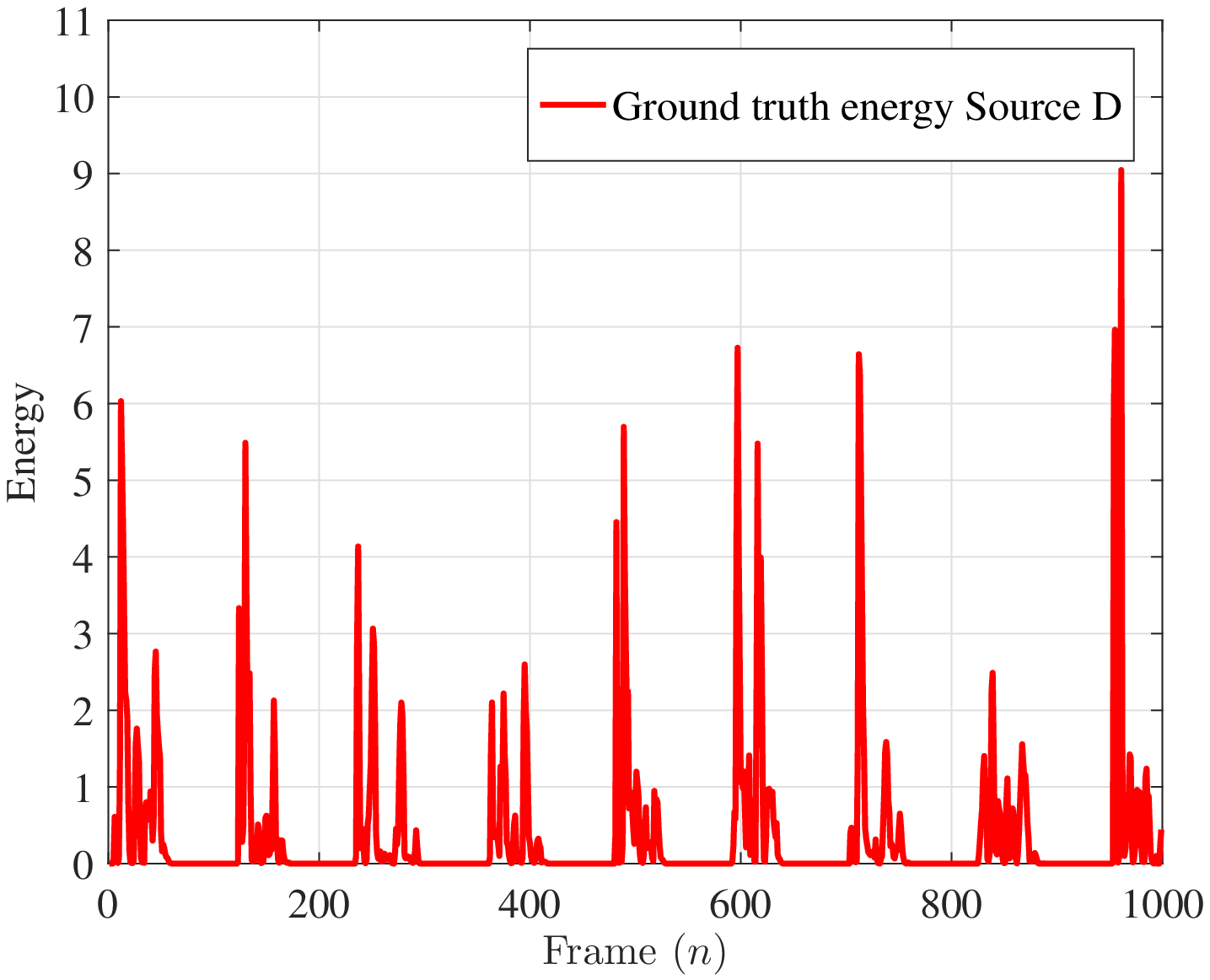}\label{fig:sub1}
 \subcaption{}
\end{minipage}%
\begin{minipage}{0.33\linewidth}
  \centering
  \includegraphics[width=46 mm, height=35mm]{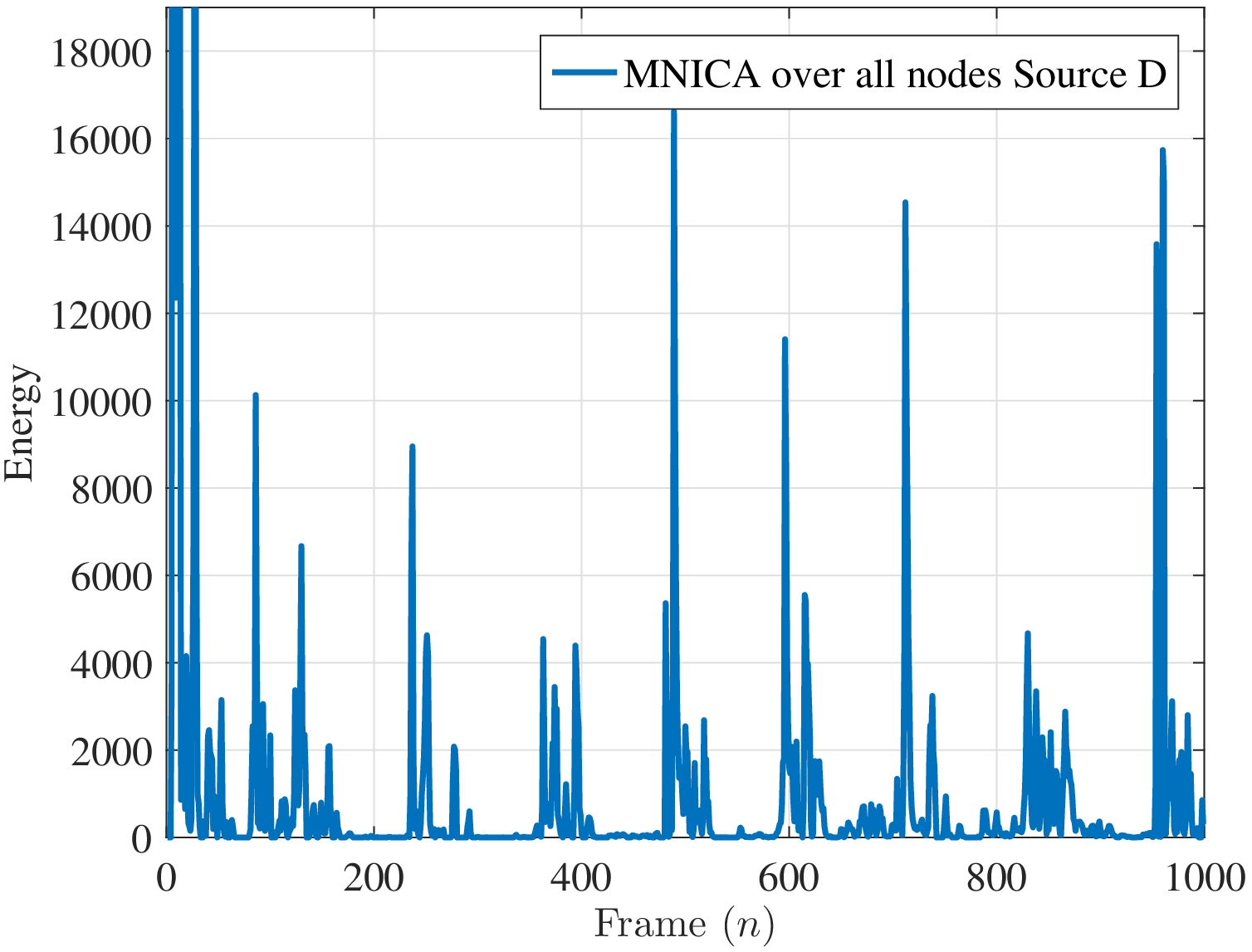}\label{fig:sub2}
  \subcaption{}
\end{minipage}
\begin{minipage}{0.33\linewidth}
  \centering
  \includegraphics[width=46 mm, height=35mm]{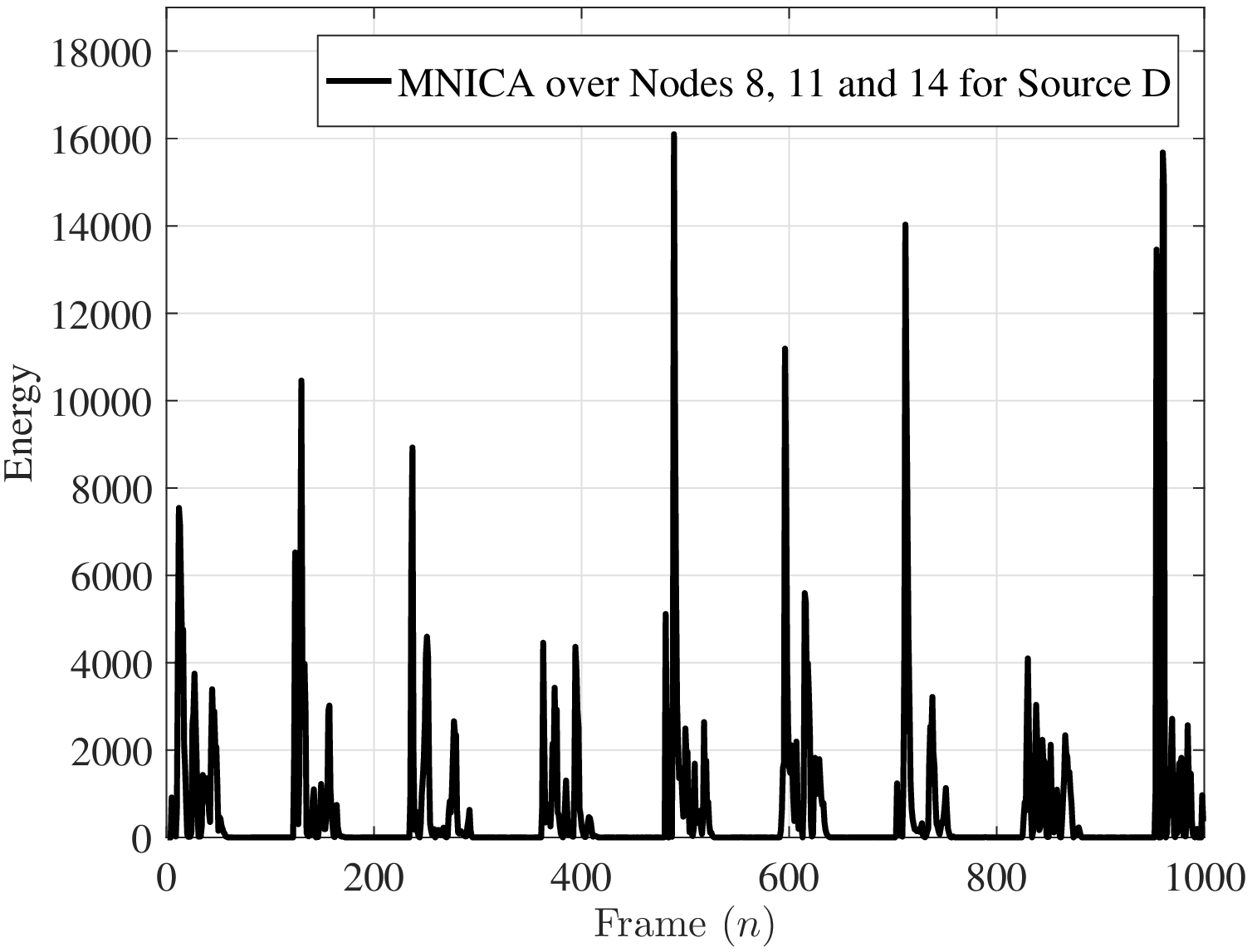}\label{fig:sub3}
\subcaption{}
\end{minipage}
\caption{The unmixing result for Source D in the scenario of Fig.~\ref{fig:scenario} using (b) MNICA over the observation of all nodes and (c) MNICA with the observations of Nodes 8, 11 and 14.}
\label{fig:UNMIXING_nodes_D}
\end{figure*}

Different centralized non-negative signal unmixing methods have been suggested in the literature, for example, non-negative principal component analysis (NPCA) \cite{oja2003blind} and multiplicative non-negative independent component analysis (MNICA) \cite{bertrand2010blind}. They can be used to obtain separated source energy signals, from the nodes' observations~\cite{bertrand2010energy}. However, these methods require a fusion center and furthermore their unmixing performance severely degrades with an increasing number of active sources, see for instance Fig.~\ref{fig:UNMIXING_nodes_D} (b) for an example of unmixing the energy of Source D for the WASN with seven active speech sources that is displayed in Fig.~\ref{fig:scenario}.

To arrive at a DM-VAD, we suggest to exploit the WASN topology: obviously, the nodes that are located in the proximity of a source observe the corresponding source signal with a higher power compared to other interfering source signals. Therefore, it is easier to unmix the energy signal of this specific source at these nodes. Fig.~\ref{fig:UNMIXING_nodes_D} (c) demonstrates the improved performance, of using MNICA with only the observations of nodes 8, 11 and 14, which are located around Source D, compared to using centralized MNICA in Fig.~\ref{fig:UNMIXING_nodes_D} (b). 

Although the above described idea is simple and promising, there are two major challenges for its practical implementation: (i) it requires a distributed method to enumerate the sources that are sensed by the WASN; (ii) it requires a distributed method to locate the nodes around each source. The first task, i.e., source enumeration, is a classical signal processing problem and can be addressed based on computing the eigenvalues of a network-wide covariance matrix~\cite{bertrand2014distributed}. The second task can be viewed as a node clustering problem, where the nodes around each source are grouped into a cluster (one cluster for each source) and the remaining nodes, which are not near to any source are grouped into an extra cluster.  In this paper, we suggest a source-specific node clustering method to locate the nodes around each source (LONAS), which yields a unified framework to solve both the source enumeration and the node clustering problem based on adaptive distributed eigenvalue 
decomposition (EVD) \cite{scaglione2008decentralized,li2010decentralized,
li2011distributed, le2008distributed,gastpar2006distributed, bertrand2014distributed}. To determine source-specific voice activity, partitional clustering algorithms are applied based on low-dimensional features that are extracted from the unmixed source energy signals.

The paper is organized as follows. Section~\ref{sec:signal-model} provides the signal model, while Section~\ref{sec:DM-VAD} details the proposed DM-VAD. Section~\ref{sec:validation1} reports on numerical experiments to evaluate the proposed method in different practical scenarios. Conclusions are drawn in Section~\ref{sec:conclusions}.

\section{Signal Model}\label{sec:signal-model}
\label{sec:problem}

A WASN of $K$ nodes whose topology is described by a graph with nodes indexed by $k\in\mathcal{K} \triangleq \{ 1,\ldots,K\}$ is considered. Node $k$ has $m_k$ microphones indexed by $m\in \{1,\ldots,M_k\}$. Its neighborhood, denoted as $\mathcal{N}_k$, is the set of nodes, including node $k$, that node $k$ exchanges information with. There are $Q$ speech sources to be detected that are indexed by $q\in\mathcal{Q} \triangleq  \{1,\ldots,Q\}$ and it is assumed that $Q<K$.

Let $y_{k,m}^{(n,f)}\in \mathbb{C}$ be a (short-term) stationary and ergodic random variable representing the observation of node $k$ at microphone $m$ and let $(n,f)$ be a point in the short-term Fourier domain, with $n \in \{1,\ldots,N\}$ indexing time segments and $f \in \{1,2,\ldots,F\}$ indexing frequency bins. Then,
\begin{eqnarray}
\label{eq:model}
y_{k,m}^{(n,f)}\approx\sum_{q=1}^{Q} h_{k,m,q}^{(n,f)}\ s_q^{(n,f)} + \varepsilon_{k,m}^{(n,f)},\end{eqnarray}
where $s_q^{(n,f)}$ refers to source signal $q$, $h_{k,m,q}^{(n,f)} \in \mathbb{C}$ is the transfer function from the $q$th source to the $m$th microphone of node $k$, and $\varepsilon_{k,m}^{(n,f)}$ is spatially uncorrelated white Gaussian noise with $\mathbb{E}|\varepsilon_{k,m}^{(n,f)}|^2 =\sigma_{k}^2$. If source $q$ is not observed by node $k$, $h_{k,m,q}^{(n,f)}=0$ $\forall (n,f)$. As a result, the nodes may observe different sets $\mathcal{Q}_k \subset \mathcal{Q}$ of sources, i.e., node $k\in\mathcal{K}$ observes $Q_k\leq Q$ sources depending on its position. 

\section{Proposed Distributed Multi-Speaker Voice Activity Detection (DM-VAD)}\label{sec:DM-VAD}

\subsection{Overview of the proposed algorithm}
\begin{figure}[h!]
	\centering
	\includegraphics[width=100mm]{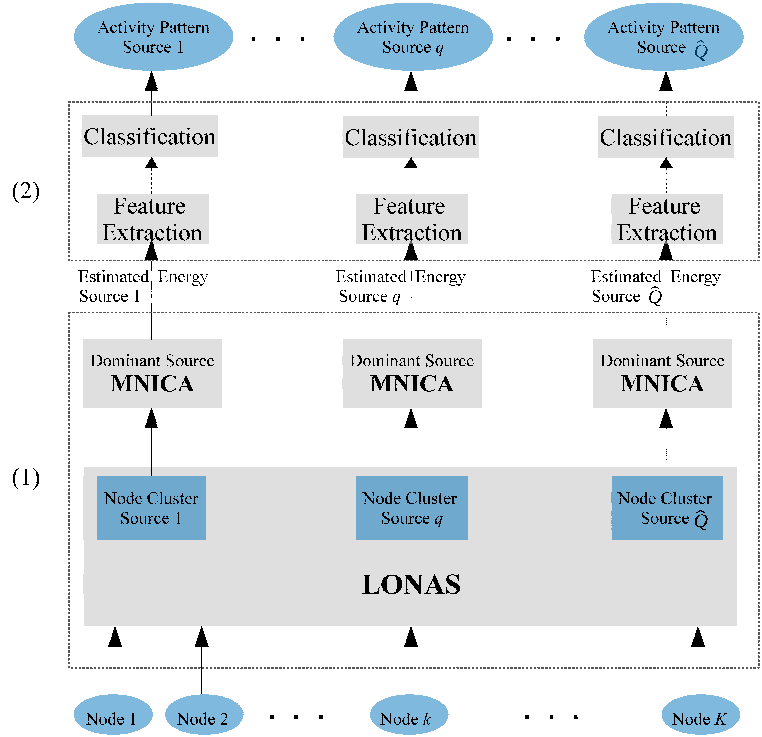}
	\caption[Two Sources.]{Block-diagram of the proposed DM-VAD.}    
	\label{Fig:DM-VAD}
\end{figure}

Figure~\ref{Fig:DM-VAD} provides an overview of the proposed DM-VAD algorithm, which consists of two main steps 

\begin{enumerate}
 \item[(1)] distributed source-specific energy signal unmixing,
 \item[(2)] energy signal based voice activity detection. 
\end{enumerate}

To unmix the source energies, a distributed method (LONAS) is proposed to identify $Q$ node clusters $\mathcal{C}_q, q\in \mathcal{Q}$, which are composed such that $\mathcal{C}_q$ observes source $q$ as the dominant speech source. LONAS requires a distributed source enumeration method to obtain an estimate of $Q$, which we denote as $\hat{Q}$. Subsequently, each cluster applies MNICA to separate its dominant source's energy from the remaining signal and noise content\footnote{Note that estimating the energy signal of a single-source using the observations of the nodes around it is a much easier task for MNICA compared to estimating $Q$ energy signals simultaneously given the observations of all the nodes in the WASN. Furthermore, scalability for large $Q$ and $K$ is obtained by the proposed divide-and-conquer strategy.}. To determine voice activity, partitional clustering algorithms are applied for which the features are extracted from the unmixed source energies to distinguish the pause from the active 
speech frames for each source 
separately. 

\subsection{{Locating Nodes Around the Sources (LONAS)}}\label{sec:signal-model_Locating}
Let $\mathbf{y}^{(n,f)} \in \mathbb{C}^{M \times 1}$, with $M=\sum_{k=1}^K M_k$, be formed from $\mathbf{y}_{k}^{(n,f)}\triangleq [y_{k,1}^{(n,f)},\ldots,y_{k,M_k}^{(n,f)}]^{\top}\in \mathbb{C}^{M_k \times 1}$ by stacking the nodes observations as follows
\begin{equation}
\label{eq:stochasticmatrices}
\mathbf{y}^{(n,f)}\triangleq \begin{bmatrix}
\mathbf{y}_{1}^{(n,f)}\\ 
\vdots\\ 
\mathbf{y}_{K}^{(n,f)}
\end{bmatrix}
\end{equation}
and let $\mathbf{Y}^{(f)}\in \mathbb{C}^{M \times N_{\mathrm{seg}}}$ be defined by
\begin{equation}
\label{eq:augmentedobservations}
\mathbf{Y}^{(f)}\triangleq\left[\mathbf{y}^{(1,f)},\ldots,\mathbf{y}^{(N_{\mathrm{seg}},f)} \right],
\end{equation}
where $N_{\mathrm{seg}}$ is the total number of --possibly-- overlapping short time Fourier transform (STFT) segments. 

By exploiting the fact that the nodes in the proximity of the same source observe similar signals, we cluster the WASN based on the node observations, i.e., the rows of observation matrix $\mathbf{Y}^{(f)}$ into $Q^+=Q+1$ clusters. We propose a new method to do this, based on the distributed EVD of the network-wide covariance matrix which is briefly described as follows.

\subsubsection{Distributed EVD}
\label{se:DEVD1}

From \eqref{eq:augmentedobservations}, the network-wide covariance matrix at the $f$th frequency bin can be defined by
\begin{equation}
\mathbf{R}^{(f)}_{YY}=E\{\mathbf{Y}^{(f)}(\mathbf{Y}^{(f)})^H\},
\end{equation}
where $E\{\cdot\}$ denotes the expected value operator, and the superscript $H$ denotes the conjugate transpose operator.

A consistent estimator of $\mathbf{R}^{(f)}_{YY}$ is the sample covariance matrix
\begin{equation}
\label{Rapprox}
\hat{\mathbf{R}}^{(f)}_{YY}= \frac{1}{N_{seg}}\mathbf{Y}^{(f)}(\mathbf{Y}^{(f)})^H.
\end{equation}

\par
The EVD of $\hat{\mathbf{R}}^{(f)}_{YY}$ is given by
\begin{equation}
\label{EVD}
\hat{\mathbf{R}}^{(f)}_{YY}=\mathbf{U}^{(f)} \boldsymbol\Sigma^{(f)}(\mathbf{U}^{(f)})^H,
\end{equation}
where $\boldsymbol\Sigma^{(f)}=\text{diag}(\lambda_1^{(f)},\hdots,\lambda^{(f)}_K)$ is a real diagonal matrix of the eigenvalues sorted in a descending order and $\mathbf{U}^{(f)}$ is a unitary matrix of the corresponding (normalized) eigenvectors in the columns. If we assume full connectivity between the nodes, $\hat{\mathbf{R}}^{(f)}_{YY}$ is easily constructed. However, in this research, we assume that each node $k\in \mathcal{K}$ is connected only within $\mathcal{N}_k$. 

The problem of distributed EVD of the network-wide covariance matrix $\hat{\mathbf{R}}^{(f)}_{YY}$ has been addressed in \cite{scaglione2008decentralized,li2010decentralized,li2011distributed, le2008distributed,gastpar2006distributed}.  
The methods in~\cite{le2008distributed,gastpar2006distributed}, require prior knowledge of the network-wide covariance matrix, while in this research the network-wide covariance matrix is assumed to be unknown and possibly even time-varying. The suggested methods in~\cite{scaglione2008decentralized,li2010decentralized,li2011distributed}, apply nested consensus averaging (CA) iterations relying on Oja's learning rule. 
The inner loop performs many CA iterations with a full reset for each outer
loop iteration running with the same rate
as the sampling rate of observation. In these methods, the convergence rate of the inner CA
loop and per-node communication cost increases with the network size.
In this research, we use the algorithm that we have recently proposed in~\cite{bertrand2014distributed}, which does 
not require nested loops, and its per-node communication cost is independent of the network size. 
The applied method, referred to as DACMEE, does not explicitly rely on Oja's stochastic learning rule and works in WSN with a tree topology. 
Unlike~\cite{scaglione2008decentralized,li2010decentralized,li2011distributed}, the applied method also estimates the eigenvectors corresponding to the $\Phi$ largest or $\Phi$ smallest eigenvalues.

DACMEE considers the following truncated versions of $\mathbf{U}^{(f)}$
\begin{equation}
\label{optsol}
\tilde{\mathbf{X}}^{(f)}=\mathbf{U}^{(f)}\left[\begin{array}{c}
\mathbf{I}_\Phi\\
\mathbf{0}_{(M-\Phi)\times \Phi}
\end{array}\right],
\end{equation}
where $\mathbf{I}_\Phi$ denotes the $\Phi\times \Phi$ identity matrix and $\mathbf{0}_{(M-\Phi)\times \Phi}$ denotes the $(M-\Phi)\times \Phi$ all-zero matrix. The rows of $\tilde{\mathbf{X}}^{(f)}$, which are applied to observations of node $k$, are denoted as $\tilde{\mathbf{X}}_{k}^{(f)}$, i.e., $(\tilde{\mathbf{X}}^{(f)})^H\hat{\mathbf{Y}}^{(f)}=\sum_{k\in\mathcal{K}}(\tilde{\mathbf{X}}^{(f)}_k)^H\hat{\mathbf{Y}}^{(f)}_k$.
It is noted that $\tilde{\mathbf{X}}$ is also a solution of the following constrained optimization problem
\begin{align}
\label{optprob}
\tilde{\mathbf{X}}^{(f)}\in\underset{\mathbf{X}^{(f)}}{\arg\max}&\: \text{Tr}\left\{(\mathbf{X}^{(f)})^H\hat{\mathbf{R}}^{(f)}_{YY}\mathbf{X}^{(f)}\right\}\\
\text{s.t.}\:\:\:&(\mathbf{X}^{(f)})^H\mathbf{X}^{(f)}=\mathbf{I}_\Phi,
\label{con}
\end{align}
where $\text{Tr}\left\{.\right\}$ denotes the trace operator.
The DACMEE algorithm~\cite{bertrand2014distributed} suggests a distributed method to solve this problem in a WASN with tree topology using an alternating optimization (AO) procedure that gradually increases the value of the objective function (\ref{optprob}) under the orthogonality constraint (\ref{con}) by updating $\hat{\mathbf{R}}^{(f)}_{k}$ and $(\bar{\mathbf{X}}_k^{(f)})$ in each node iteratively. It is proved in \cite{bertrand2014distributed} that the DACMEE procedure converges to the optimal solution of~(\ref{optprob}) yielding the eigenvectors of the network-wide covariance matrix $\hat{\mathbf{R}}^{(f)}_{YY}$.

\subsubsection{Source enumeration}
The problem of source enumeration has been widely studied. Existing methods may be loosely grouped into two categories which are associated with the hypothesis testing approach, e.g., \cite{williams1990using, brcich2002detection} and the approach based on information theoretic criteria \cite{stoica2004model}. Often, source enumeration methods are motivated by the following asymptotical property of the eigenvalues of $\hat{\mathbf{R}}_{\mathbf{YY}}$
\begin{align}
	\label{eq:GLM}
	\lambda_1\geqslant \ldots \geqslant \lambda_Q > \lambda_{Q+1}=\ldots =\lambda_M=\sigma^2,
\end{align}
where the first $Q$ eigenvalues correspond to the sources and the remaining $M-Q$ eigenvalues correspond to the spatially and temporally white noise, i.e., $\sigma_{k}^2=\sigma^2\ \forall k\in\mathcal{K}$. However, for a finite number of observations, as explained, e.g., in~\cite{lu2013generalized}
\begin{align}
	\label{eq:GLM}
	\lambda_1>\ldots>\lambda_Q>\lambda_{Q+1}>\ldots >\lambda_M.
\end{align}

In the following, a source enumeration method using a Lasso penalized Poisson (LAPPO) generalized linear model is proposed to identify source-related eigenvalues of $\hat{\mathbf{R}}_{\mathbf{YY}}$. The generalized linear model with a Lasso penalization is often used in model selection, since in this method predictors with negligible explanatory properties are shrunken to zero~\cite{tibshirani1996regression,mccullagh1989generalized,friedman2001elements,tang2014risk}. 

The source eigenvalues are obtained as non-zero elements of a vector $\boldsymbol{\beta}=[\beta_1,\ldots, \beta_M]^{\top}$ through the following minimization problem
\begin{equation}
\label{eq:lossfunction}
\underset{\boldsymbol{\beta},\beta_0}{\min}\ \frac{1}{2 N_w} \sum^{N_w}_{i=1}\left(\ln(z_i)-\beta_0-\mathbf{w}_i \boldsymbol{\beta} \right)^2+\alpha \sum^{M}_{m=1} |\beta_m|, 
\end{equation}
where $\beta_0$ is a sequence offset, $\alpha$ is a hyperparameter that tunes the contribution of the Lasso penalization, $z_i$ follows a Poisson model with log-link as link function, i.e., 
\begin{equation}
\label{eq:link}
z_i\sim \mathrm{Po}(\tilde{z}_i)\quad \mathrm{with} \ \tilde{z}_i=\mathrm{exp}\left(\frac{\mathbf{w}_i\boldsymbol{\lambda}}{\sum_{m=1}^M \lambda_m}\right),
\end{equation}
$\boldsymbol{\lambda}\triangleq(\lambda_1,\ldots,\lambda_M)^\top$, $\mathbf{w}_i\triangleq(w_{i1},\ldots,w_{iM}) \in \mathbb{R}_0^+$ is a weighting vector. An arbitrarily large number of weighting vectors $\mathbf{w}_i$, e.g., $N_w=1000$ is generated randomly from a multinomial distribution and a 10-fold cross-validation approach is used to tune $\alpha$. Different methods have been suggested to solve \eqref{eq:lossfunction}, e.g., active set, block coordinate~\cite{banerjee2008model,friedman2008sparse} and grafting~\cite{lee2006efficient}. In this paper, the spectral projected gradient method~\cite{schmidt2009optimizing,bertsekas1982projected} is used. The number of non-zero elements of $\boldsymbol{\beta}$ is the estimated number of sources $\hat{Q}$.

\textbf{Remark 1}: To keep the communication cost low, instead of calculating all eigenvalues/eigenvectors of $\hat{\mathbf{R}}^{(f)}_{YY}$, we compute the $\Phi$ largest eigenvalues/eigenvectors using the DACMEE algorithm. In this case, the parameter $\Phi$ should be determined such that it is larger than $\hat{Q}$. 

 \textbf{Remark 2}: The DACMEE algorithm yields the eigenvectors of the network-wide covariance matrix $\hat{\mathbf{R}}^{(f)}_{YY}$ for each frequency bin $f$.  Therefore, LAPPO yields $\hat{Q}$ for each frequency bin, so we can take the average between the obtained estimation results to reach a single estimate of $\hat{Q}$.  
 
\subsubsection{Distributed source-specific node clustering}
Since the nodes around a source observe similar signals, clustering the nodes' observations results in locating the nodes around the sources.
In this task, the rows of $\mathbf{Y}^{(f)}$ are clustered into $\hat{Q}^{+}$ clusters, one cluster for each source and an additional cluster for the remaining nodes. Different distributed clustering methods can be adopted for this task, e.g. the distributed K-means algorithm~\cite{jagannathan2005privacy} and the distributed Gaussian mixture model~\cite{gu2008distributed}. However, these methods require many iterations of consensus averaging over high-dimensional observation vectors, which is energy inefficient.
To avoid this, we use the method of Zha {\it et al.}~\cite{zha2001spectral} and Ding {\it et al.}~\cite{ding2004k}, and effectively make use of the estimated eigenvalues $\boldsymbol{\Sigma}$ and eigenvectors $\mathbf{U}$.

Using the results of \cite{zha2001spectral} and \cite{ding2004k}, the continuous cluster membership indicator matrix $\tilde{\mathbf{B}}$ of dimension $M \times \hat{Q}$ can be obtained through the optimization problem
\begin{align}
\label{eq:zha}
{\tilde{\mathbf{B}}}^{(f)}\in\underset{{\mathbf{B}}^{(f)}}{\arg\min}&\:  \text{Tr} \left[({\mathbf{B}}^{(f)})^{\sf{H}}\hat{\mathbf{R}}^{(f)}_{\mathbf{YY}}{\mathbf{B}^{(f)}}\right],
\end{align}
s.t.
\begin{align}
\label{eq:con_zha}
({\tilde{\mathbf{B}}}^{(f)})^{\sf{H}}{\tilde{\mathbf{B}}}^{(f)}=\mathbf{I}_{{\hat{Q}}},
\end{align}
where $\mathbf{I}_{\hat{Q}}$ is the $\hat{Q}$-dimensional identity matrix. The clustering optimization problem of~(\ref{eq:zha})-(\ref{eq:con_zha}) is identical to the EVD problem \eqref{EVD}, hence no extra calculation is required and the continuous cluster membership indicator matrix ${\tilde{\mathbf{B}}}^{(f)}$ corresponds to the first ${{\hat{Q}}}$ eigenvectors of $\hat{\mathbf{R}}^{(f)}_{\mathbf{YY}}$ contained in the full matrix of eigenvectors ${\mathbf{B}}^{(f)} \in \mathbb{C}^{M \times M}$. Similar to LAPPO the clustering can be performed locally at each node since the eigenvectors are known in all the nodes after applying the DACMEE algorithm and sharing the results.  

\textbf{Remark 3}: Note that ${\tilde{\mathbf{B}}}^{(f)}$ is complex-valued. To perform the clustering we use its magnitude only, which is denoted as $|{\tilde{\mathbf{B}}}^{(f)}|$.

\textbf{Remark 4}: Note that $|{\tilde{\mathbf{B}}}^{(f)}|$ is calculated for each frequency bin through the DACMEE algorithm. Therefore, to reach a single clustering result, we take an average, i.e. $|{\tilde{\mathbf{B}}}|=\sum_{f}|{\tilde{\mathbf{B}}}^{(f)}|$.

A computationally simple and accurate method to obtain the clusters from $|{\tilde{\mathbf{B}}}|$ is the normalized cut (\textrm{NCut}) algorithm~\cite{shi2000normalized}, which has been successfully used for image segmentation and spectral clustering~\cite{shi1998image,shi2000normalized,gao2016graph}. In this method, the data is represented as a graph $G\left(\mathcal{B},\mathcal{E}\right)$. The set of vertices $\mathcal{B}$ is the set of data points $\mathbf{b}_{i}$ for $i\in\{1,\ldots,M \}$, where $\mathbf{b}_{i}$ is the ${i}$-th row of $|{\tilde{\mathbf{B}}}|$. $\mathcal{E}$ denotes the set of edges between the data points $\mathbf{b}_{i}$ and $\mathbf{b}_{j}$ with weights $\theta_{i,j}>0$, which represent a similarity measure between the two data points. The affinity matrix of $G\left(\mathcal{B},\mathcal{E}\right)$ is the matrix $\boldsymbol{\Theta}$ whose elements are $\theta_{i,j}$ with $i\in\{1,\ldots,M \}$ and $j\in\{1,\ldots,\hat{Q} \}$. 

The NCut algorithm separates $G\left(\mathcal{B},\mathcal{E}\right)$ into two disjoint sets $\mathcal{Z}$ and $\bar{\mathcal{Z}}$ by solving
\begin{align}
\label{eq:ncut}
\underset{\boldsymbol{\omega}}{\min}\ \frac{\boldsymbol{\omega}^{\top}\left(\mathbf{D}-\boldsymbol{\Theta}\right)\boldsymbol{\omega}}{\boldsymbol{\omega}^{\top}\mathbf{D}\boldsymbol{\omega}}
\end{align}
s.t. 
\begin{equation}
\label{eq:ncut_discrete}
\omega_m\in \{-\frac{\mathrm{Vol}\left(\mathcal{Z}\right)}{\mathrm{Vol}\left(\bar{\mathcal{Z}}\right)},1\}
\end{equation}
and
\begin{equation}
\label{eq:ncut_unity}
 \boldsymbol{\omega} \mathbf{D} \mathbf{1}_{M}=0.
\end{equation}
Here, $\boldsymbol{\omega}=(\omega_1,\ldots,\omega_M)^\top$ denotes the class indicator vector, $\mathbf{D}\in\mathbb{R}^{M\times M}$ is a diagonal matrix with elements
\begin{equation}
 d_{mm} \triangleq \sum_{j=1}^{\hat{Q}}\theta_{mj}
\end{equation}
and
\begin{equation}
 \mathrm{Vol}\left(\mathcal{Z}\right) \triangleq  \sum_{m\in \mathcal{Z}} d_{mm}. 
\end{equation}
Unfortunately, the optimization problem (\ref{eq:ncut})-(\ref{eq:ncut_unity}) is NP-complete. However, an approximate
solution can be found if $\omega_m$ is relaxed to take real values. Its approximate solution, namely the Fiedler vector~\cite{shi2000normalized}, follows from the second smallest eigenvector of the matrix $\left(\mathbf{D}-\boldsymbol{\Theta}\right)$ such that nodes corresponding to non-negative components of the Fiedler vector are clustered in $\mathcal{Z}$ and the remaining nodes are clustered in $\bar{\mathcal{Z}}$. Next, the cluster with the lowest algebraic connectivity is bi-partitioned to obtain three clusters. This process is repeated until $\hat{Q}^{+}$ disjoint clusters of nodes are obtained.

\subsection{Distributed unmixing of the source energy signals}\label{sec:signal-model_Feature}

The instantaneous energy of source signal $q$ at time segment $n$ equals
\begin{align}
\label{eq:energy}
\bar{s}_q^{(n)}=\frac{1}{F}\sum_{f=1}^{F} |s_q^{(n,f)}|^2.
\end{align}
Similarly, the instantaneous energy of microphone signal $m$ at node $k$ in time segment $n$ is given by
\begin{align}
\label{eq:en}
\bar{y}_{k,m}^{(n)}=\frac{1}{F}\sum_{f=1}^{F} |y_{k,m}^{(n,f)}|^2.
\end{align}
Assuming that the source signals are mutually independent and neglecting the reverberation effects over time segments \cite{bertrand2010energy}, we then have
\begin{align}
\label{eq:centralized}
\bar{\mathbf{y}}^{(n)}\approx \mathbf{H}\bar{\mathbf{s}}^{(n)},
\end{align}
with
\begin{align}
& \bar{\mathbf{y}}^{(n)}\triangleq\left[(\bar{\mathbf{y}}_1^{(n)})^{\top},\ldots, (\bar{\mathbf{y}}_k^{(n)})^{\top}, \ldots, (\bar{\mathbf{y}}_K^{(n)})^{\top}\right]^{\top}\label{eq:energy_network}\\
&\bar{\mathbf{y}}^{(n)}_k\triangleq\left[\bar{y}_{k,1}^{(n)},\ldots, \bar{y}_{k,M_k}^{(n)}\right]^{\top}\label{eq:energy_nodes}\\
&\bar{\mathbf{s}}^{(n)}\triangleq\left[\bar{s}_1^{(n)},\ldots, \bar{s}_Q^{(n)}\right]^{\top},\label{eq:energy_sources}
\end{align}
where $\mathbf{H}\in\mathbb{R}^{M\times Q}$ is a mixing matrix that describes the power attenuation between the sources and the microphones.

Centralized unmixing of the energies based on \eqref{eq:centralized} can be performed for example, using non-negative principal component analysis (NPCA) \cite{oja2003blind} and multiplicative non-negative independent component analysis (MNICA) \cite{bertrand2010blind,bertrand2010energy}. As exemplified in Fig.~\ref{fig:UNMIXING_nodes_D} (b), even with the availability of a fusion center, the performance severely degrades for increasing values of $Q$. In order to overcome these restraints, we propose a decentralized LONAS based unmixing approach. As mentioned earlier, $\mathcal{C}_q\subset \mathcal{K}$ denote the set of nodes that are assigned to the $q$th source by LONAS, and let $|\mathcal{C}_q|>0$ denote its cardinality, i.e., the number of nodes assigned to the $q$th source. Further, analogously to \eqref{eq:energy_network}, let $\bar{\mathbf{y}}_{\mathcal{C}_q}^{(n)}\in\mathbb{C}^{|\mathcal{C}_q|\times 1}$ contain the instantaneous energies of the microphone signals of all nodes $k\in\mathcal{C}_q$ 
at time segment $n$. Then, 
assuming that $\bar{s}_q^{(n)}$ is the dominant source for the nodes in $\mathcal{C}_q$ we define
\begin{align}
\label{eq:source_q}
\bar{\mathbf{y}}_{\mathcal{C}_q}^{(n)}\approx \mathbf{h}_{\mathcal{C}_q}\bar{s}_q^{(n)}, \quad q\in \{1,\ldots, \hat{Q}\},
\end{align}
where $\mathbf{h}_{\mathcal{C}_q}$ is a $|\mathcal{C}_q|$-dimensional mixing vector that describes the power attenuation between the $q$th source and the nodes within $\mathcal{C}_q$. Based on \eqref{eq:source_q}, each cluster $\mathcal{C}_q, q\in \{1,\ldots, \hat{Q}\}$ uses a source-specific (two-component\footnote{which the first output is the unmixed energy of the nearest source and the second output is the energy of background noise (the effect of other sources).}) MNICA algorithm to determine $\bar{s}_q$. This means that for the proposed distributed approach $\hat{Q}$ two-component MNICA algorithms are used,  instead of a single centralized MNICA that assumes $\hat{Q}$ sources. For this reason, in the proposed approach, the performance of the energy unmixing no longer depends on $\hat{Q}$.

\subsection{Distributed dominant speaker VAD}
\label{sec:detection}
The final step of the proposed algorithm distinguishes the active and the non-active speech segments for each source by means of efficient partitional clustering algorithms \cite{Theodoridis:2008:PRF:1457541,theodoridis2014machine}. These determine the class membership of each time segment, depending on its distance to the estimated cluster centroids.

Let $\bar{s}_{\mathcal{C}_q}^{(n)},\quad q=1,\ldots, \hat{Q}$ denote the estimates of the source-specific energy signals $\bar{s}_q^{(n)}$. Source-specific voice activity patterns for each source $q=1,\ldots, \hat{Q}$ are determined by extracting features from $\bar{s}_{\mathcal{C}_q}^{(n)}$ locally within each node cluster $\mathcal{C}_q$, allowing for a distributed computation\footnote{Unique labels of $\bar{s}_{\mathcal{C}_q}^{(n)}$ throughout the network are available from the distributed labelling algorithm presented in \cite{chouvardas2015distributed}.}. The feature vector
\begin{equation}\label{eq:4}
 \boldsymbol{v}_q^{(n)}\triangleq [v_{q,1}^{(n)}, v_{q,2}^{(n)}, v_{q,3}^{(n)}]^{\top} 
\end{equation}
is formed from the following features\footnote{The feature selection is the result of an empirical study that contained a larger set of features which we do not elaborate on for the sake of conciseness.} 
\begin{enumerate} 
\item the short-term arithmetic average
\begin{equation}\label{eq:1}
 v_{q,1}^{(n)}\!=\!\frac{1}{W} \sum_{i=n-W}^{n + 1}\!\!\!\! \bar{s}_{\mathcal{C}_q}^{(i)}, \quad n\in\{W+1,\cdots,N\}
\end{equation}
\item the short-term standard deviation
 \begin{equation}\label{eq:2}
 v_{q,2}^{(n)}\!=\!\sqrt{\frac{1}{W} \sum_{i=n-W}^{n}\!\!\!\!  ( \bar{s}_{\mathcal{C}_q}^{(i)}-v_{q,1}^{(i)})^2}, \quad n\in\{W+1,\cdots,N\}
\end{equation}
\item the first-order energy difference
\begin{equation}\label{eq:3}
 v_{q,3}^{(n)}= \bar{s}_{\mathcal{C}_q}^{(n)}-\bar{s}_{\mathcal{C}_q}^{(n+1)}, \quad n\in\{W,\cdots,N-1\}
\end{equation}
\end{enumerate}
\begin{figure}[h!]
\centering
\includegraphics[width=100mm]{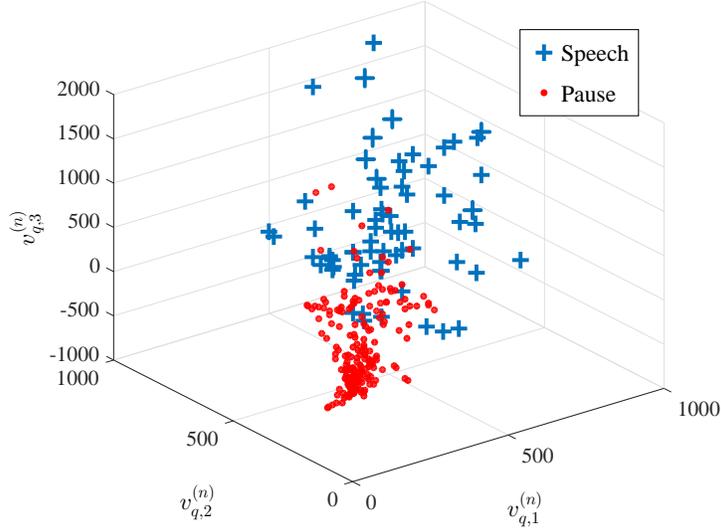}
\caption[Data spread after clustering.]{Example of the extracted feature vectors for Source A.}    
\label{fig:fig5}
\end{figure}
Figure~\ref{fig:fig5} gives an illustrating example, where each point corresponds to one feature vector $ \boldsymbol{v}_q^{(n)}$, which either belongs to the active speech (blue crosses) or the pause class (red dots). Since the distribution of the data in the feature space is non-Gaussian, robust variations of the classic K-means algorithm are considered. One such variation is the K-medians, where the sample median of each cluster is used to determine its centroid. This results in minimizing the error over all clusters with respect to the $\ell_1$-norm distance metric. A further variation is the K-medoids, which can be used with arbitrary metrics of distances, and is based on the medoid, which is the instance from the dataset for which the average dissimilarity to all the objects in the cluster is minimal.

Let ${\hat{\mathbf{c}}}_{j}^{(q)}\triangleq [\hat{c}_{j,1}^{(q)},\hat{c}_{j,2}^{(q)},\hat{c}_{j,3}^{(q)}]^\top\in\mathbb{R}^{3\times 1}, \forall j \in {1,2}$ denote the estimated centroids of the K-means type algorithms for source $q\in\mathcal{Q}$, and let ${\hat{\mathbf{c}}}_1^{(q)}$ correspond to the pause class, which is easily identified by $\min~\hat{c}_{j,1}^{(q)}$ for $j\in\{1,2\}$ since the short-term average energy of this class is smaller than for the active speech class, and then ${\hat{\mathbf{c}}}_1^{(q)}$ corresponds to the active speech class.

The cluster memberships are determined from
\begin{equation}
\label{eq:15}
   {t}_{j}(\boldsymbol{v}_q^{(n)}) =\lVert \boldsymbol{v}_q^{(n)} - {\hat{\mathbf{c}}}_{j}^{(q)} \rVert _2^2 , \quad n \in \{W+1,\dots,N\},
\end{equation}
based on which a decision is formed by

\begin{equation}
\label{eq:17}
 \delta_q^{(n)} = \left\lbrace 
\begin{tabular}{c c c}
 0 & if ${t}_1(\boldsymbol{v}_q^{(n)}) < {t}_2(\boldsymbol{v}_q^{(n)})$ & (pause), \\
 1 & otherwise & (active speech). \\
\end{tabular} 
\right.
\end{equation}

\subsubsection{Remark}
\label{subsec:practical-remark}
\begin{figure}[h!]
\centering
\includegraphics[width=100mm]{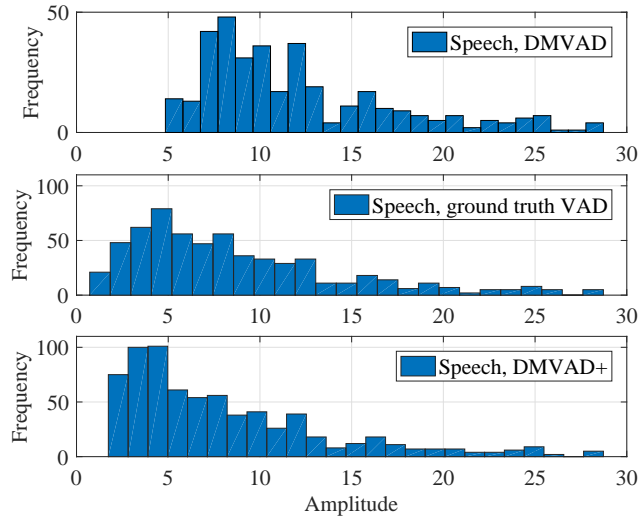}
\caption[improving misdetection rate.]{Example of the histogram of $v_{q,1}^{(n)},\ n\in\{W+1,\ldots,N\}$ for Source A for the active speech class using the distributed multi-speaker VAD (DMVAD) approach before (top), i.e. DMVAD, and after (bottom), i.e. DMVAD+, applying the correction step defined in Eq. (\ref{eq:20}), and the ground truth histogram for Source A in the middle.}    
\label{fig:fig1}
\end{figure}

Because of its practical usefulness, we report on a simple and optional correction step. This step aims at reducing the misdetection rate by reassigning the labels for some low power data points that were falsely assigned to the pause class in $\delta_q^{(n)}$ by the decision rule defined in Eq.~(\ref{eq:17}). Given the assignments from (\ref{eq:17}), let $\mathcal{M}_0$ denote the set of $n\in\{W+1,\ldots,N\}$ for which $\delta_q^{(n)} =0$ and let $\mathcal{M}_1$ denote the set where $\delta_q^{(n)} =1$. Then with
\begin{equation*}
\hat{\sigma}_0\triangleq \mathrm{mad}(\{v_{q,1}^{(n)}\} ), \quad \forall n\in\mathcal{M}_0
\end{equation*}
\begin{equation*}
\hat{\sigma}_1\triangleq \mathrm{mad}(\{v_{q,1}^{(n)}\} ), \quad \forall n\in\mathcal{M}_1
\end{equation*}
where $\mathrm{mad}(\mathcal{X})$ is the median absolute deviation of a dataset $\mathcal{X}$. By defining
\begin{equation}
\label{eq:18}
  D_0(n) \triangleq \lvert v_{q,1}^{(n)} - \hat{\sigma}_0 \rvert 
\end{equation}
and
\begin{equation}
\label{eq:19}
  D_1(n) \triangleq \lvert v_{q,1}^{(n)} - \hat{\sigma}_1 \rvert 
\end{equation}
the voice activity decision $\delta_{q}^{(n)}$ may be corrected as follows: 
\begin{equation}
\label{eq:20}
  \delta_{q,\textrm{new}}^{(n)} = \left\lbrace 
 \begin{tabular}{c c}
  $1$ & if $D_0(n) > D_1(n), $ \\
  $\delta_q^{(n)}$ & otherwise.
 \end{tabular} 
 \right.
\end{equation}
Figure~\ref{fig:fig1} illustrates the effect of the correction step on the empirical distribution function of the energies $v_{q,1}^{(n)}$ that are associated to the speech class.
In the top graph, the histogram based on the assignments of Eq.~(\ref{eq:17}) is displayed. The middle graph shows the distribution of speech obtained for the case of the (unavailable) ground truth assignments. The lower graph depicts the speech distribution after applying the correction step of Eq.~(\ref{eq:20}). It is noticed that the speech distribution after the correction step becomes more similar to the one obtained from the ground truth assignments.
A shift in the speech distribution mode to the left is observed in the bottom subplot. This is explained by the correct reassignement of elements to the speech distribution. 
The positive effect of this correction step is also noticed in the real data experiments, see Tabs.~\ref{table:table11}-\ref{table:table14} and Tab.~\ref{table:table20}.
\subsection{Batch-Mode and Sequential VAD Algorithm}
\label{sec:step5}
The proposed VAD algorithm, which is run locally, e.g., by a unique node at each node cluster $\mathcal{C}_q, q\in\mathcal{Q}$ can be operated on batches of data (batch-mode VAD), or for streaming data (sequential VAD). The batch mode VAD algorithm is summarized in Algorithm~\ref{table:table1}. 
\begin{algorithm}[h]
\caption{Batch-Mode VAD algorithm for Source $q$ evaluated locally within $\mathcal{C}_q$.}
\label{table:table1}
\begin{algorithmic}[1]
\Statex \textbf{Input}
\State Set a value for $W$
\Statex \textbf{Batch VAD procedure}
\For{$n=W+1,\ldots,N$} 
\State Compute the features $\boldsymbol{v}_q^{(n)}$ using (\ref{eq:1})-(\ref{eq:4}).
\EndFor
\State Estimate the centroids $\hat{\mathbf{c}}_{j}^{(q)}, j=1,2$ using
\Statex  K-means, K-medians or K-medoids.
\State Label $\min(\hat{c}_{j,1}^{(q)}), j\in\{1,2\}$ and $\max(\hat{c}_{j,1}^{(q)}), j\in\{1,2\}$
\Statex as pause and active speech centroids, respectively.
\State Decide $\forall n\in \{W+1,\ldots,N\} $ based on $\eqref{eq:17}$
\Statex \textbf{Output}
\State VAD patterns in $\delta_q^{(n)}, \forall n\in \{W+1,\ldots,N\}$.
\end{algorithmic}
\end{algorithm}
In the sequential VAD algorithm, the feature vector $\boldsymbol{v}_q^{(n)}$ is calculated sequentially for streaming-in unmixed energy signals $\bar{s}_{\mathcal{C}_q}^{(n)}, n=W+1,\ldots$ which can be computed with the adaptive MNICA algorithm as described in \cite{bertrand2010energy}. In the sequential mode, the proposed VAD algorithm uses a growing window so as to incorporate all past information\footnote{In principle, a sliding window implementation is also possible, however, the window must be chosen large enough so as to capture both active speech and pause segments.}. 
In this case, the instantaneous feature vectors are obtained by evaluating (\ref{eq:1}), (\ref{eq:2}), and (\ref{eq:3}) for all time segments $n$. The features at each time segment $n$ are collected as in (\ref{eq:4}).  All further steps are the same as in the batch mode algorithm, given the available data, except that the random initialization of the centroids in the sequential VAD algorithm is performed only once. Then the sequential VAD uses the previous value of the centroid estimates as initialization. 

After operating Algorithm~\ref{table:table1}, the extracted VAD patterns are shared within and between clusters. 
\section{Validation}
\label{sec:validation1}
Numerical experiments are conducted to assess the performance of our proposed DM-VAD. All proposed subsystems are evaluated individually and compared to existing benchmarks, wherever possible. 

\subsection{Validation of Lasso penalized Poisson (LAPPO) source enumeration}
\label{sec:Eval_lse}
\begin{figure}[h]
	\centering
	\includegraphics[width=100 mm]{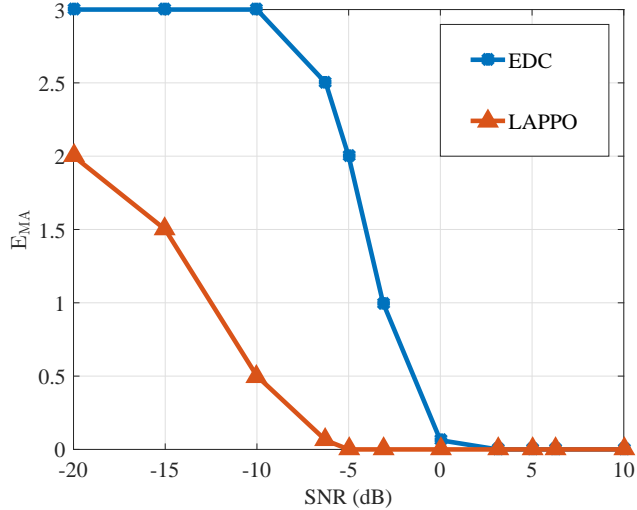}
	\caption{The $\mathrm{E_{MA}}$ for different values of observation SNR.}    
	\label{fig:lappo}
\end{figure}

The accuracy of LAPPO source enumeration for speaker enumeration is validated via an acoustic scenario which is generated using the image method~\cite{Habets,allen:79}. A three dimensional rectangular room $(10 \mathrm{m} \times 20 \mathrm{m} \times 6 \mathrm{m})$ with reflection coefficients of $0.3$ at all walls is considered. The WASN consists of $K=20$ nodes, each equipped with a uniform linear array of $M_k=3$ microphones with an inter-microphone distance of $1.5\mathrm{cm}$ and sampling frequency of $16\mathrm{kHz}$. $Q=4$ speech sources located at random positions, each speaking different sentences, are present. Spatially uncorrelated AWGN of equal variance $\sigma_k^2$ is superimposed at each microphone signal. To estimate the number of sources, a Hann-windowed DFT of size $2048$ with $50\%$ overlap is applied. The performance is measured by the mean absolute error ($\mathrm{E_{MA}}$) of source enumeration averaged over 20 experiments. The proposed criterion is benchmarked against the efficient 
detection criterion (EDC)~\cite{zhao1986detection}\footnote{Classical information criteria, such as MDL and Akaike fails in this setup, therefore, their performance is not reported upon.}, which is provided with the same eigenvalues. Figure~\ref{fig:lappo} displays $\mathrm{E_{MA}}$ for different values of the observation SNR for an observation length of 10 seconds. LAPPO  outperforms EDC, especially for SNR$<0$ dB. The experiment has been repeated for 1-7 active speech sources and in all experiments a similar pattern to the one shown in Fig.~\ref{fig:lappo} was obtained for LAPPO, which confirms the effectiveness of the proposed source enumeration method.  

\subsection{Validation of distributed source-specific node clustering (LONAS)}
\label{sec:Eval_lonas}
The effectiveness of the LONAS method in locating the nodes around the sources is investigated over a generic scenarios with synthetic data and over the multi-speaker WASN scenario presented in Fig.~\ref{fig:scenario}.

\subsubsection{Synthetic data scenario} LONAS is benchmarked against the sparsity-aware matrix decomposition (SMD) algorithm~\cite{6542703} and its distributed version D-SMD. We adopt the simulation setting of~\cite{6542703}, where a WASN of 12 nodes and 3 sources is considered. Source 1 is observed by Nodes 4, 5, 6 and 7, Source 2 is observed by Nodes 8, 9 and 10, and Source 3 is observed by Nodes 1 and 3. The remaining nodes observe none of the sources. The resulting nine non-zero entries of the transfer function matrix are randomly selected from the Gaussian distribution with mean one and variances $5\times10^{-4}$, $10^{-3}$ and $10^{-3}$ for entries corresponding to Sources 1, 2 and 3, respectively. The experiment is repeated for $N_{\mathrm{trl}}=500$ Monte Carlo runs. The performance is measured by the classification error rate ($\mathrm{E_{mis}}$) in (\%) 
\begin{align}\label{eq:Emis}
\mathrm{E_{mis}}=\frac{\sum_{j=1}^{N_{\mathrm{trl}}}{N^{(j)}_{\mathrm{mis}}}}{K N_{\mathrm{trl}}}\times 100,
\end{align}
where $N^{(j)}_{\mathrm{mis}}$ is the number of incorrectly classified nodes in the $j${th} trial.

\begin{figure}[h!]
	\centering
	\includegraphics[width=100 mm]{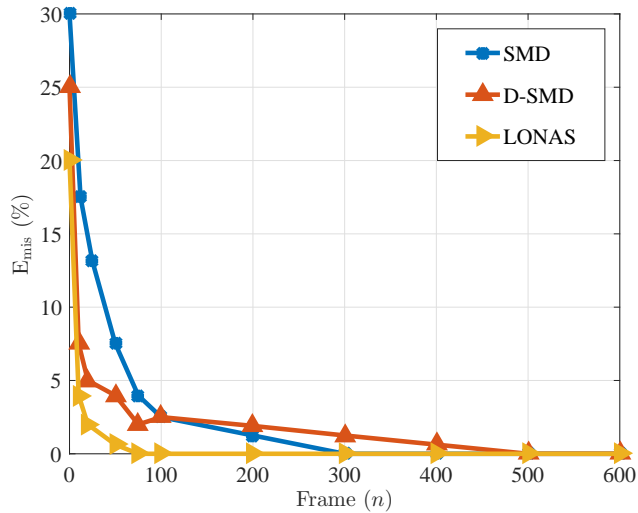}
	\caption{The $\mathrm{E_{mis}}$ for different numbers of training samples.}    
	\label{fig:schizas_test}
\end{figure}

\begin{figure}[h!]
	\centering
	\includegraphics[width=100 mm]{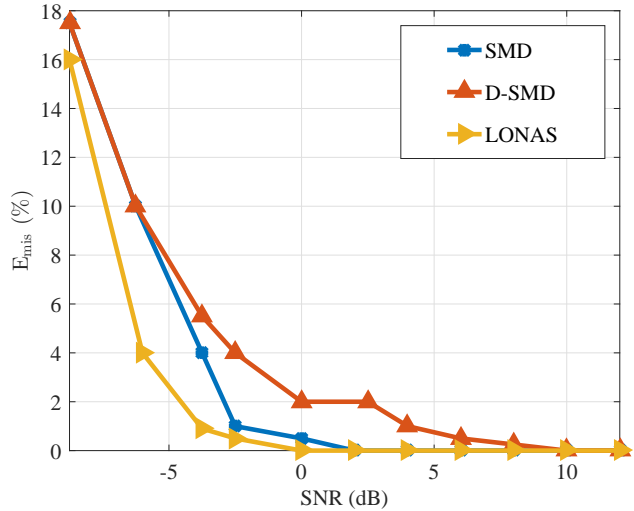}
	\caption{The $\mathrm{E_{mis}}$ for different values of the observation SNR.}    
	\label{fig:schizas_test_2}
\end{figure}

Fig.~\ref{fig:schizas_test} displays the error rate $\mathrm{E_{mis}}$ for different numbers of training samples for SNR = 18 dB. As the training data increases, $\mathrm{E_{mis}}\rightarrow 0$. Further, LONAS is shown to be more accurate than both SMD and D-SMD. Fig.~\ref{fig:schizas_test_2} shows the error rate $\mathrm{E_{mis}}$ for different values of the observation SNR, where the number of training samples is fixed to 200. Again, LONAS is more accurate than SMD and D-SMD in locating the nodes around the sources.\newline

\subsubsection{Validation over speech data}
As a realistic scenario, the WASN diplayed in Fig.~\ref{fig:scenario} is considered. Table~\ref{tab:scenario_results} lists the resulting clusters of using LONAS for an observed signal length of 60 seconds and a frame-length and frame-shift of 1024 and 512 samples, respectively. The clustering using LONAS can effectively locate the nodes around each source. Interestingly, for the public address system of source A, nodes 2 and 12 are selected, which are both near a PA loudspeaker while being far from each other spatially.   

\begin{table} [h]
\normalsize
\centerline{
\begin{tabular}{cc}
\toprule
Source    & Nodes Forming a Cluster \\
\midrule  \midrule
A& 2 and 12 \\
\midrule
B& 5, 6 and 15 \\
\midrule
C         & 7 and 18 \\
\midrule
D& 8, 11 and 14 \\	
\midrule
E& 3, 9, 13 and 16\\	
\midrule
F& 10\\	
\midrule
Dummy Nodes   & 1, 4, 17, 19 and 20\\	
\bottomrule		
\end{tabular}}
\caption{{\it The clustering result of the proposed LONAS algorithm for the scenario described in Fig.~\ref{fig:scenario}.}}
\label{tab:scenario_results}
\end{table}
	
\subsection{Validation of the LONAS based unmixing of the source energies}
\label{sec:Eval_lonas}
\begin{figure}[t]
\begin{minipage}{0.33\linewidth}
  \centering
 \includegraphics[width=46 mm, height=35mm]{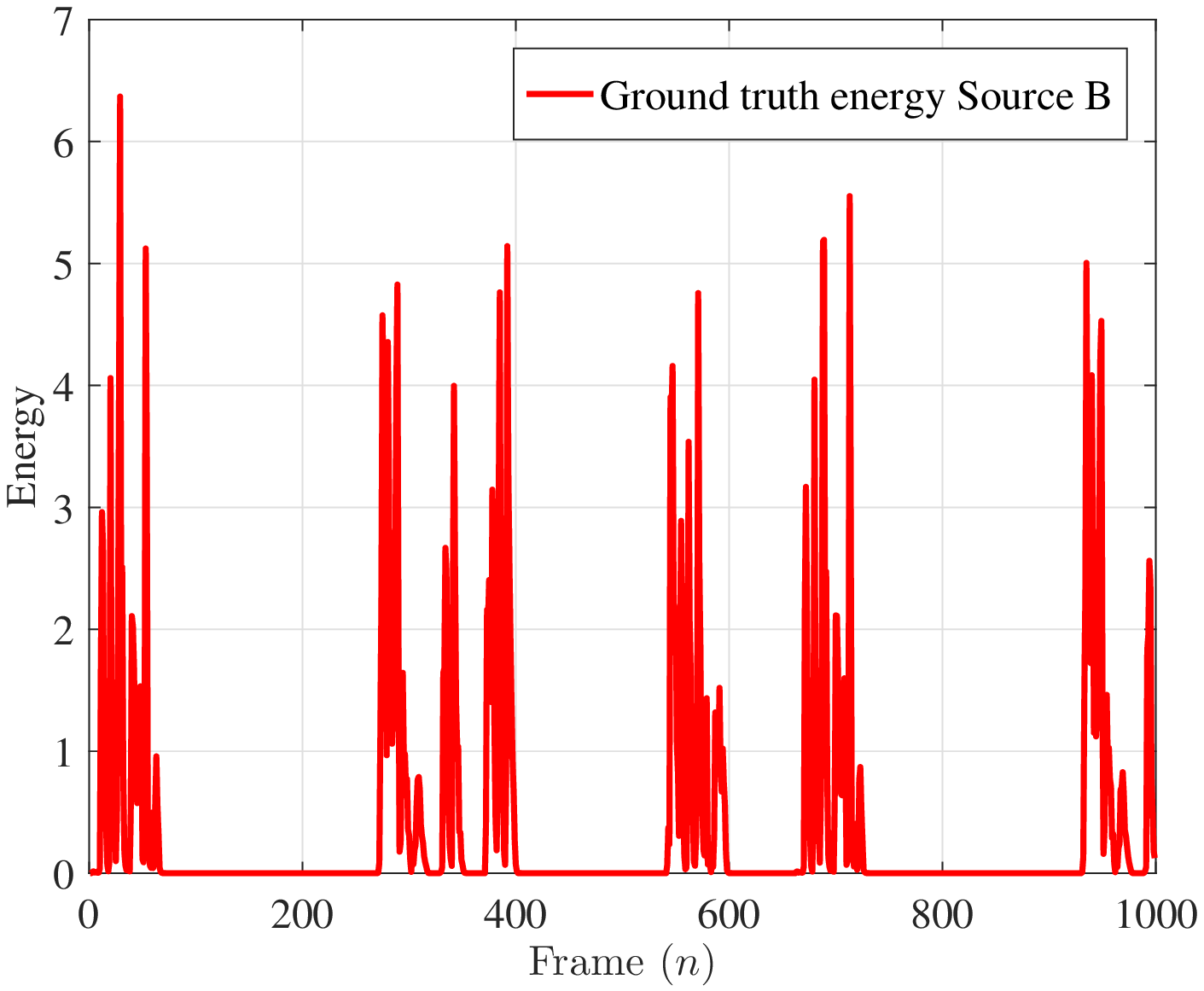}\label{fig:sub4}
 \subcaption{}
\end{minipage}%
\begin{minipage}{0.33\linewidth}
  \centering
  \includegraphics[width=46 mm, height=35mm]{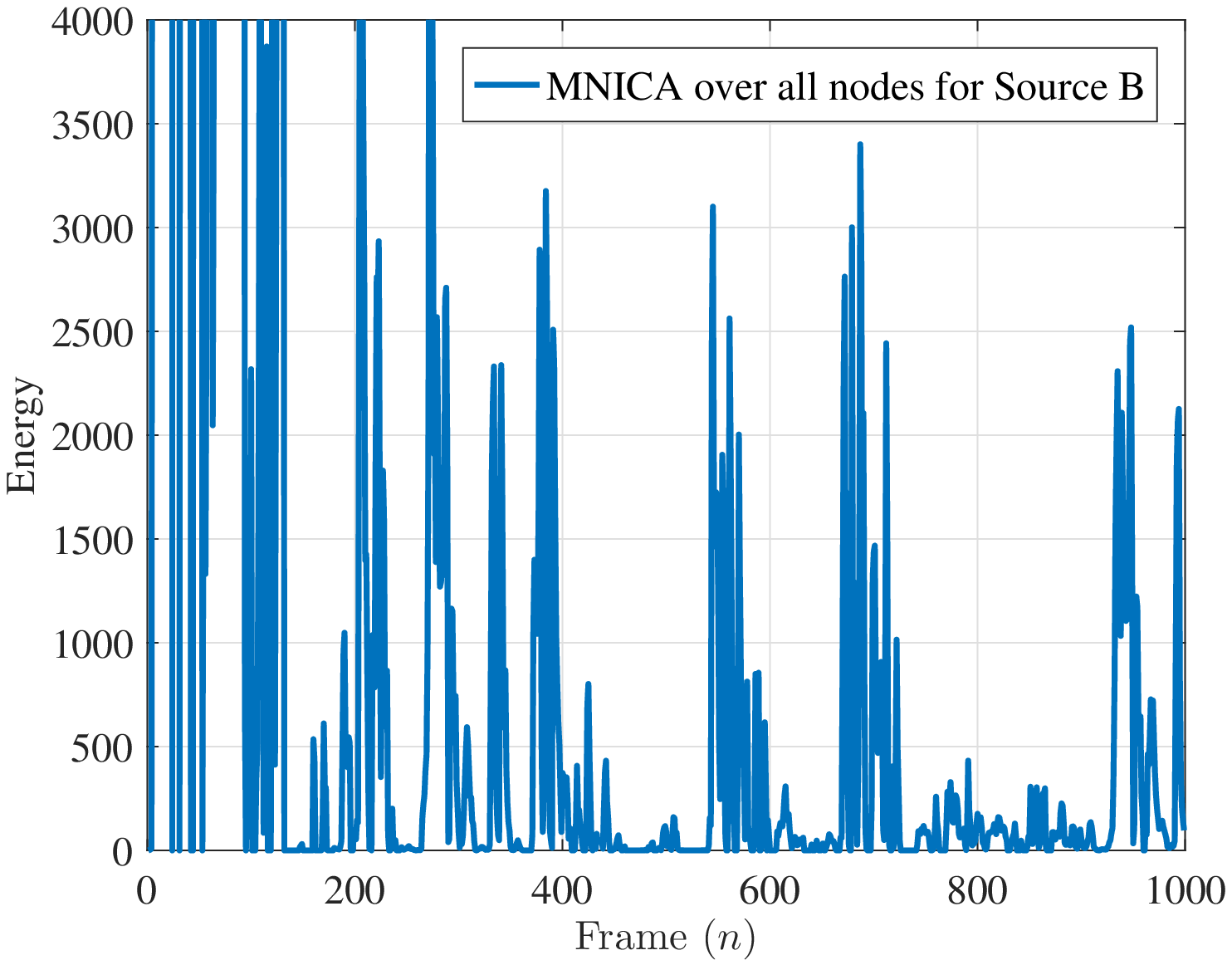}\label{fig:sub5}
  \subcaption{}
\end{minipage}
\begin{minipage}{0.33\linewidth}
  \centering
  \includegraphics[width=46 mm, height=35mm]{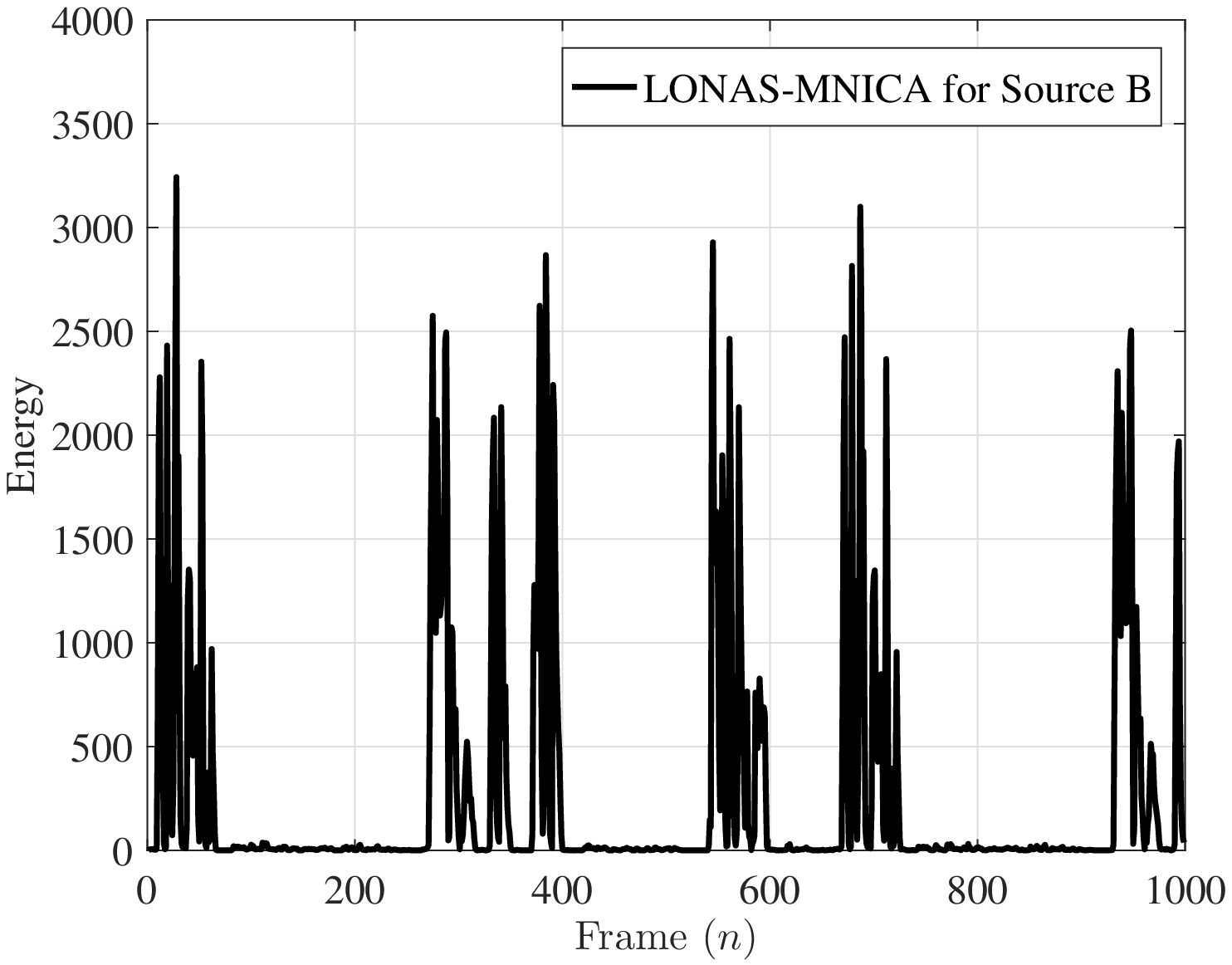}\label{fig:sub6}
\subcaption{}
\end{minipage}
\caption{The unmixing results for Source B using (b) MNICA over all nodes and (c) LONAS-MNICA.}
\label{fig:UNMIXING_nodes_B}
\end{figure}

\begin{figure}[t]
\begin{minipage}{0.33\linewidth}
  \centering
 \includegraphics[width=46 mm, height=35mm]{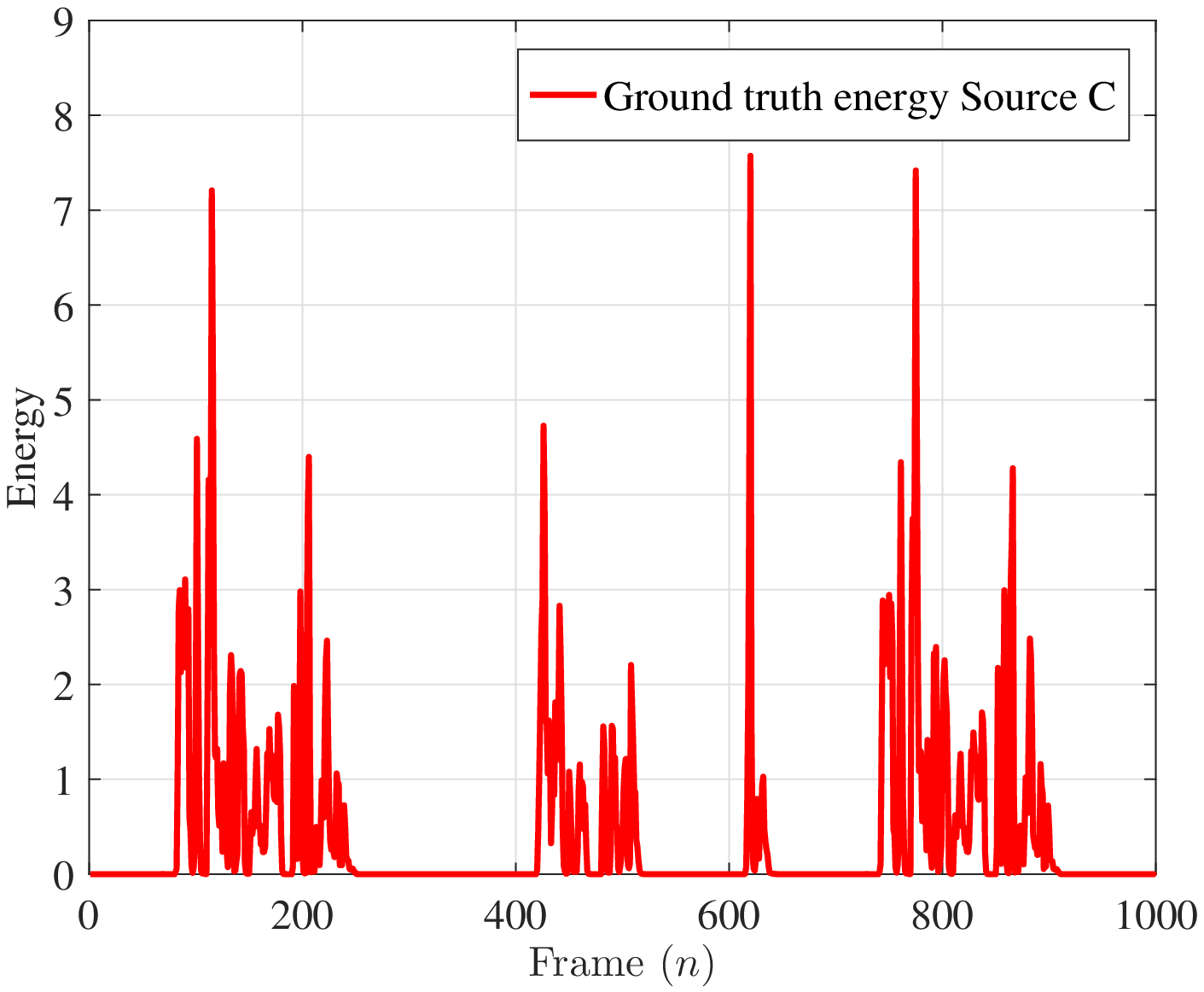}\label{fig:sub7}
 \subcaption{}
\end{minipage}%
\begin{minipage}{0.33\linewidth}
  \centering
  \includegraphics[width=46 mm, height=35mm]{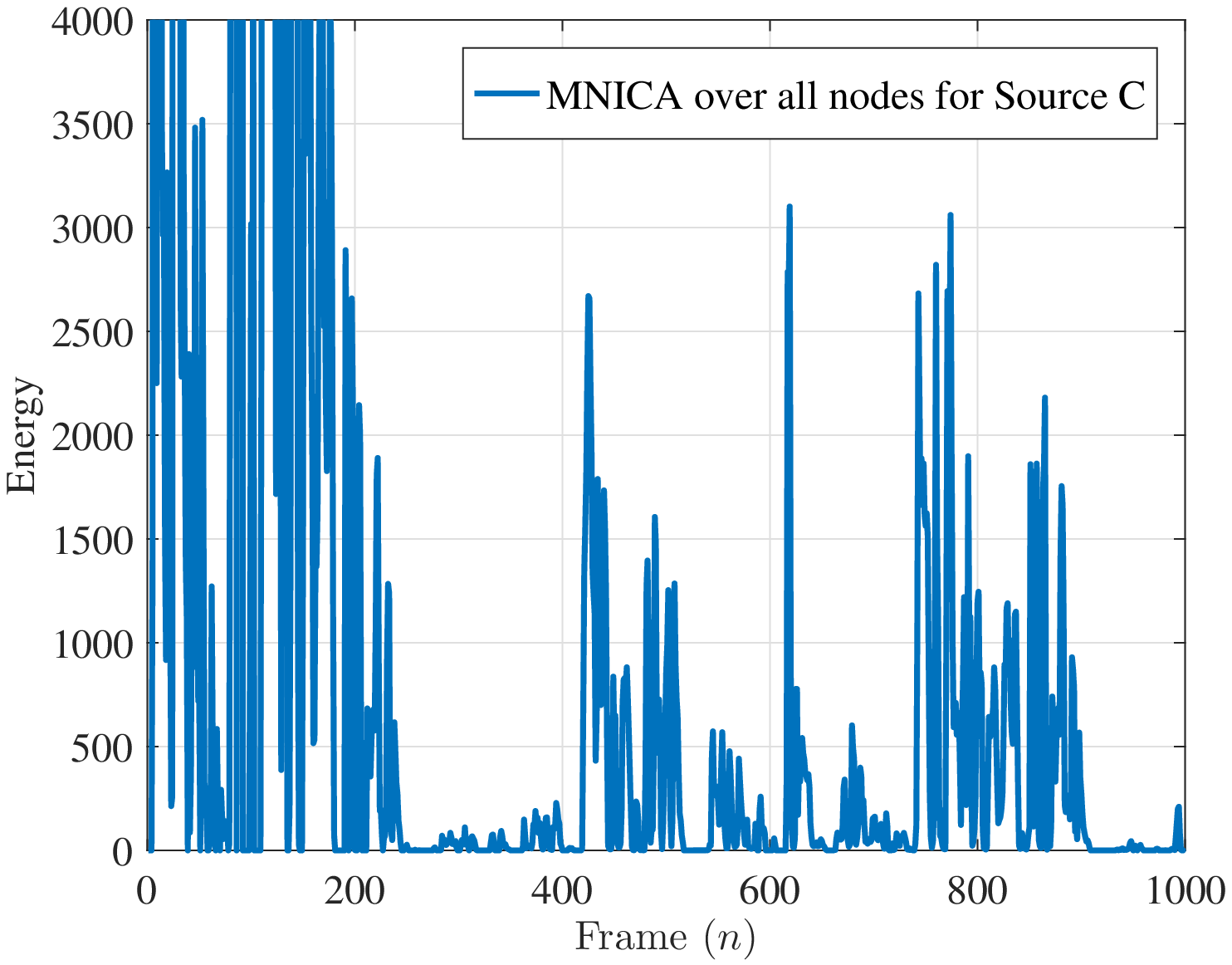}\label{fig:sub8}
\subcaption{}
\end{minipage}
\begin{minipage}{0.33\linewidth}
  \centering
  \includegraphics[width=46 mm, height=35mm]{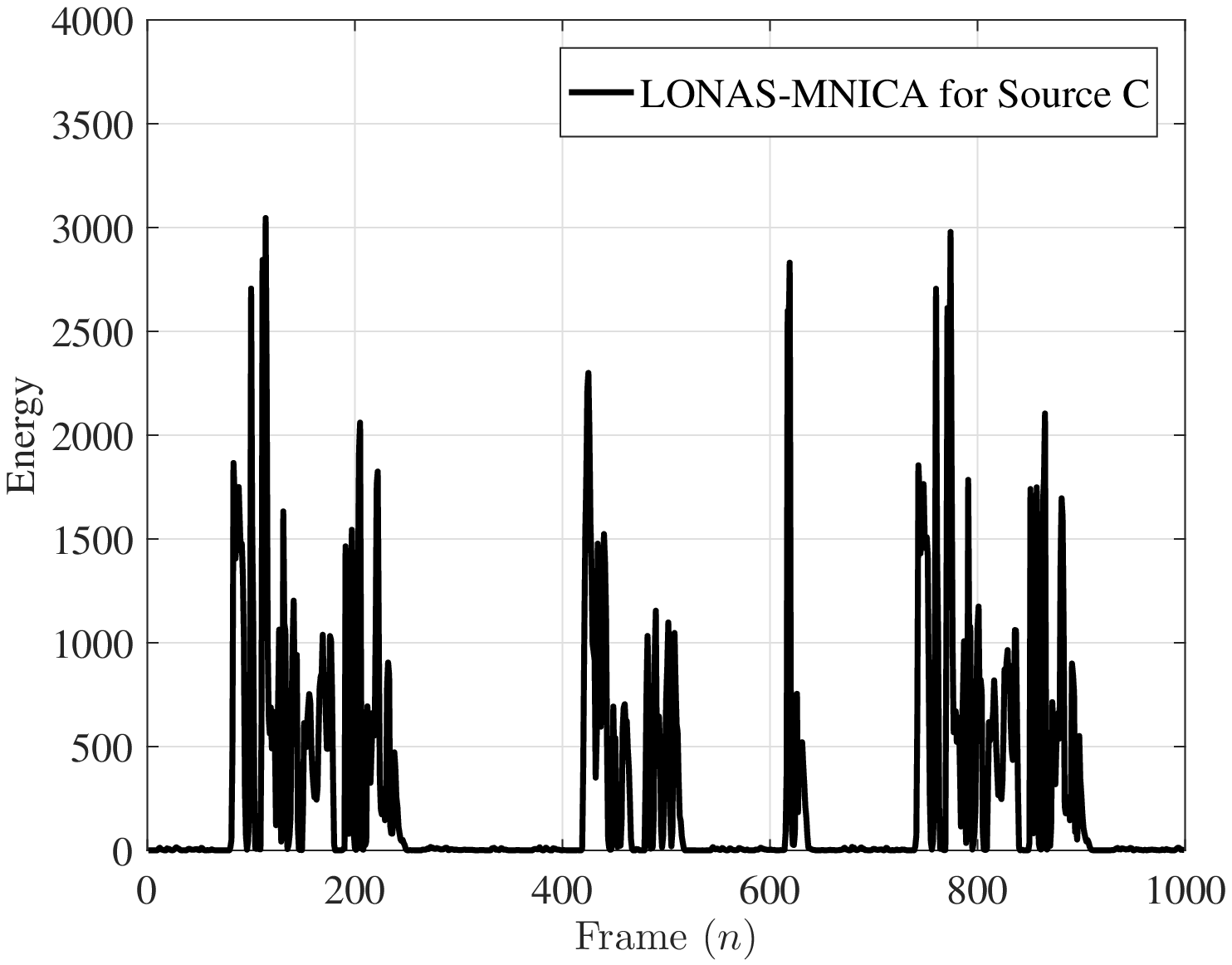}\label{fig:sub9}
  \subcaption{}
\end{minipage}
\caption{The unmixing results for Source C using (b) MNICA over all nodes and (c) LONAS-MNICA.}
\label{fig:UNMIXING_nodes_C}
\end{figure}

The results of the distributed LONAS-MNICA unmixing of the source energies (see Eq. \eqref{eq:source_q}) is benchmarked against the centralized MNICA unmixing with all nodes~\cite{bertrand2010blind} (see Eq. \eqref{eq:centralized}) using the scenario of Fig.~\ref{fig:scenario}. The performance is measured by the Pearson correlation coefficient ($\eta_{\mathrm{cc}}$) between the actual source energy and the corresponding estimate.  
Table \ref{tab:MNICA} shows that the proposed distributed LONAS-MNICA considerably outperforms its centralized counterpart. The output of MNICA with LONAS and MNICA over all nodes for Sources B and C is shown in Figs.~\ref{fig:UNMIXING_nodes_B} and~\ref{fig:UNMIXING_nodes_C}, respectively.
\begin{table} [h]
\normalsize
\centerline{
\begin{tabular}{cccc}
\toprule
Source    & All Nodes & LONAS & Relative Improvement (\%)\\
\midrule  \midrule
A & 0.83 & 0.90 & 8.5\\
\midrule
B	     & 0.73 & 0.76 & 4\\
\midrule
C         & 0.62 & 0.78 & 20\\
\midrule
D& 0.72& 0.81 & 11\\	
\midrule
E & 0.45 & 0.74 & 39\\	                                                                                
\midrule
F & 0.68& 0.72 & 4.5\\	
\bottomrule
\end{tabular}}
\caption{{\it The Pearson correlation coefficient $\eta_{\mathrm{cc}}$ for all sources using centralized MNICA over all nodes and LONAS-MNICA.}}
\label{tab:MNICA}
\end{table}

\subsection{Validation of detection}
\label{sec:Eval_detect}
In this section, the accuracy of the proposed detection method is verified for single-speaker and multi-speaker scenarios by considering the WASN displayed in Fig.~\ref{fig:scenario}.
 
\subsubsection{Single-speaker scenario}
The performance for single-speaker VAD is benchmarked against two existing single-node methods, i.e., the VAD-1 \cite{VAD3} and the VAD-2 \cite{verteletskaya2010voice} given observations from Node 2 for Source A and Node 9 for Source E, respectively. The distributed multi-speaker VAD (DMVAD) refers to the proposed VAD approach based on K-medoids and without post-processing (see Sec. \ref{subsec:practical-remark}), whereas DMVAD+ includes this step. Tables \ref{table:table11}-\ref{table:table14} summarize the results of the comparative study for Sources A and E under Gaussian and babble noise conditions of variance $\sigma^2=0.01$. In the ensuing tables, the values in bold are indicators of a superior performance attained by our proposed VAD technique. The performance metrics are: correct decision (CD), missed detection (MD), false alarm (FA), equal error rate (EER), and cost of log-likelihood ratio ($C_{\textrm{llr}}^{\min}$). The EER reports the measure between the frame-level speech and non-speech 
detections and $C_{\textrm{llr}}^{\min}$ measures the quality of the log-likelihood ratio detection output. In both cases, a small value corresponds to a highly accurate VAD. DMVAD+ outperforms its single-node competitors by 
leveraging upon the WASN via LONAS-MNICA and achieves $>92 \%$ correct VAD in all cases.

\begin{table}[H]
\centering
\normalsize
\begin{tabular}{cccccc}
\toprule
\multirow{2}{*}{Metric} &\multicolumn{4}{c}{VAD Results for Source A}\\
\cmidrule{2-5}
&DMVAD & DMVAD+ & VAD-1 & VAD-2\\
\midrule \midrule
CD & 64.8 & \bf{92.3} & 89.5 & 63 \\
\midrule
MD & 34.7 & 5.7 & 3 & 0  \\
\midrule
FA & \bf{0.5} & 2 & 7.5 & 37 \\
\midrule
EER & 0.06 & \bf{0.06} & 0.4 & 0.4\\
\midrule
$C_{\textrm{llr}}^{\min}$ & 0.2 & \bf{0.2} & 0.9 & 0.9 \\
 \bottomrule
\end{tabular}
\caption{Comparison of our approach with different benchmark algorithms, referred to as VAD-1 \cite{VAD3} and VAD-2 \cite{verteletskaya2010voice}, for a single active Source A and additive white Gaussian noise of variance $\sigma^2 =0.01$.}
\label{table:table11}
\end{table}
For the same noisy environment and a different single Source E, Table \ref{table:table12} shows that DMVAD+, VAD-1 and VAD-2 provide nearly perfect detection results.
\begin{table}[H]
\centering
\normalsize
\begin{tabular}{ccccc}
\toprule
\multirow{2}{*}{Metric} &\multicolumn{4}{c}{VAD Results for Source E}\\
\cmidrule{2-5}
& DMVAD & DMVAD+ & VAD-1 & VAD-2\\
\midrule \midrule
CD & 80.3 & 96.2 & 94.6 & 96.3 \\
\midrule
MD & 19.7 & 3.8 & 3.3 & 3.1  \\
\midrule
FA & \bf{0} & \bf{0} & 2 & 0.6 \\
\midrule
EER &0.01 & \bf{0.01} & 0.35 & 0.35\\
\midrule
$C_{\textrm{llr}}^{\min}$ & \bf{0.03} & \bf{0.04} & 0.85 & 0.85 \\
 \bottomrule
\end{tabular}
\caption{Comparison of our approach with different benchmark algorithms \cite{VAD3,verteletskaya2010voice}, for a single active source E and additive white Gaussian noise of variance $\sigma^2 =0.01$.}
\label{table:table12}
\end{table}
In Tables \ref{table:table13}~and~\ref{table:table14}, we consider the single speech sources A and E corrupted with babble noise. Results show that VAD-1 and VAD-2 are more sensitive to babble noise since they loose in detection performance while the decisions in DMVAD and DMVAD+ remain stable.
\begin{table}[H]
\centering
\normalsize
\begin{tabular}{ccccc}
\toprule
\multirow{2}{*}{Metric} &\multicolumn{4}{c}{VAD Results for Source A}\\
\cmidrule{2-5}
& DMVAD & DMVAD+ & VAD-1 & VAD-2\\
\midrule \midrule
CD & 65 & \bf{92.7} & 88.2 & 61.9 \\
\midrule
MD & 34.5 & 5.2 & 2.2 & 0  \\
\midrule
FA & \bf{0.5} & 2.1 & 29.6 & 38.1 \\
\midrule
EER & 0.06 & \bf{0.06} & 0.4 & 0.4\\
\midrule
$C_{\textrm{llr}}^{\min}$ & 0.2 & \bf{0.2} & 0.9 & 0.9 \\
 \bottomrule
\end{tabular}
\caption{Comparison of our approach with different benchmark algorithms \cite{VAD3,verteletskaya2010voice}, for a single active source A and babble noise of variance $\sigma^2 =0.01$.}
\label{table:table13}
\end{table}

\begin{table}[H]
\centering
\normalsize
\begin{tabular}{ccccc}
\toprule
 \multirow{2}{*}{Metric} &\multicolumn{4}{c}{VAD Results for Source E}\\
\cmidrule{2-5}
&DMVAD & DMVAD+ & VAD-1 & VAD-2\\
\midrule \midrule
CD& 80.3 & \bf{96.2} & 94.6 & 57.6 \\
\midrule
 MD & 19.7 & 3.8 & 3.3 & 20.8  \\
\midrule
FA& \bf{0} & \bf{0} & 2 & 21.6 \\
\midrule
EER& 0.01 & \bf{0.01} & 0.35 & 0.35\\
\midrule
$C_{\textrm{llr}}^{\min}$& 0.03 & \bf{0.04} & 0.85 & 0.85 \\
 \bottomrule 
\end{tabular}
\caption{Comparison of our approach with different benchmark algorithms \cite{VAD3,verteletskaya2010voice}, for a single active source E and babble noise of variance $\sigma^2 =0.01$.}
\label{table:table14}
\end{table}

\subsubsection{Batch-mode distributed multi-speaker voice activity detection}
The performance of the proposed detector in batch-mode (see Algorithm~\ref{table:table1}) is evaluated on the challenging multi-speaker scenario with seven active sources, as given in Fig.~\ref{fig:scenario} for different variations of K-means. Tables \ref{table:table19} and \ref{table:table20} summarize the outcome of DMVAD and DMVAD+ for AWGN of variance $\sigma^2=0.01$. Comparable detection results are achieved when alternating between the variants of the K-means algorithm, and the post processing-step is most useful for Source A, where the original speech signal is a noisy PA announcement. The worst-case CD for DMVAD+ is $>84 \%$ in this challenging scenario.

\begin{table}[H]
\centering
\footnotesize
\begin{tabular}{cccccccc}
\toprule
\multirow{2}{*}{Method} & \multirow{2}{*}{Metric} & \multicolumn{6}{c}{DMVAD}\\
\cline{3-8}
&& A & B & C & D & E & F\\
\midrule \midrule
\multirow{5}{*}{K-means} & CD & 61.1 & 92.3 & 93.6 &61.7 &86.7& 92.4\\
& MD & \bf{38.1} & 7.4 &5.5  &38.3 &13.2&6.7 \\
& FA & \bf{0.8} & 0.3 & 0.9 &\bf{0} & \bf{0} & 0.9 \\
& EER & 0.22 & \bf{0.02} & \bf{0.02} &0.17 &0.03&\bf{0.04} \\
& $C_{\textrm{llr}}^{\min}$ & 0.63 & \bf{0.09} &0.11  &0.41 &0.12&0.18 \\
\midrule
\multirow{5}{*}{K-medians} & CD & \bf{70.4} & \bf{93.1} & \bf{95.2} &\bf{86.6} & \bf{88.7} & \bf{93.8}\\
& MD & 27.5 & \bf{6.6} & \bf{3.8} & \bf{13.5} & \bf{11.3} & \bf{5} \\
& FA & 2.12 & 0.3 & 1 & \bf{0} &\bf{0} & 1.2 \\
& EER & \bf{0.14} & \bf{0.02} & \bf{0.02} & \bf{0.01}&\bf{0.01} &\bf{0.04} \\
& $C_{\textrm{llr}}^{\min}$ & 0.55 & \bf{0.09} & 0.11 & \bf{0.06} & \bf{0.03} &0.14 \\
\midrule
\multirow{5}{*}{K-medoids} & CD & 62.7 & 85 & 82.1 & 74.7&80.3&88.1 \\
& MD & 36.3 & 14.9 & 17.6 &25.3 &19.7&11.2 \\
& FA & 1 & \bf{0.1} & \bf{0.3} & \bf{0} &\bf{0} & \bf{0.7} \\
& EER & 0.15 & \bf{0.02} & \bf{0.02} & \bf{0.01}&\bf{0.01}&\bf{0.04} \\
& $C_{\textrm{llr}}^{\min}$ & \bf{0.5} & \bf{0.09} & \bf{0.1} & \bf{0.06}&\bf{0.03}& \bf{0.12} \\
\bottomrule
\end{tabular}
\caption{Proposed batch-mode DMVAD using different clustering methods for signals corrupted by additive white Gaussian noise of variance $\sigma^2=0.01$.}
\label{table:table19}
\end{table}

\begin{table}[H]
\centering
\footnotesize
\begin{tabular}{cccccccc}
\toprule
\multirow{2}{*}{Method} & \multirow{2}{*}{Metric} & \multicolumn{6}{c}{DMVAD+}\\
\cmidrule{3-8}
&& A & B & C & D & E & F\\
\midrule \midrule
\multirow{5}{*}{K-means} & CD & 85.2 & 96.2 & 97 &89.9 & \bf{96.2} & \bf{94.8}\\
& MD & 5.1 & \bf{0.8} &0.9  &10.1 & \bf{3.8} & \bf{2.2} \\
& FA & 9.7 & 3 & \bf{2.1} & \bf{0} & \bf{0} & \bf{2.9} \\
& EER & \bf{0.15} & \bf{0.02} & \bf{0.02} &\bf{0.01} &0.03& \bf{0.04} \\
& $C_{\textrm{llr}}^{\min}$ & 0.5 & \bf{0.09} & \bf{0.1}  &0.07 &0.12&0.18 \\
\midrule
\multirow{5}{*}{K-medians} & CD & \bf{86.3} & \bf{96.3} & 97 &\bf{93.6} &\bf{96.2}& \bf{94.8}\\
&  MD & \bf{3.5} & \bf{0.8} & 0.9 & \bf{6.4} & \bf{3.8} & \bf{2.2} \\
& FA & 10.1 & \bf{2.9} & \bf{2.1} & \bf{0} & \bf{0} & \bf{2.9} \\
& EER & \bf{0.15} & \bf{0.02} & \bf{0.02} & \bf{0.01} &\bf{0.01} & \bf{0.04} \\
& $C_{\textrm{llr}}^{\min}$ & \bf{0.49} & \bf{0.09} & 0.11 & \bf{0.06} & \bf{0.04} & \bf{0.14} \\
\midrule
\multirow{5}{*}{K-medoids} & CD & 84.6 & \bf{96.3} & \bf{97.1} & \bf{93.6}& \bf{96.2} & \bf{94.8} \\
& MD & 6 & \bf{0.8} & \bf{0.8} &\bf{6.4} & \bf{3.8} & \bf{2.2} \\
& FA & \bf{9.4} & \bf{2.9} & \bf{2.1} & \bf{0} & \bf{0} & \bf{2.9} \\
& EER & \bf{0.15} & \bf{0.02} & \bf{0.02} & \bf{0.01} & \bf{0.01} & \bf{0.04} \\
& $C_{\textrm{llr}}^{\min}$ & 0.5 & \bf{0.09} & \bf{0.1} & \bf{0.06} & \bf{0.04} & \bf{0.14} \\
\bottomrule
\end{tabular}
\caption{Proposed batch-mode DMVAD+ using different clustering methods for signals corrupted by additive white Gaussian noise of variance $\sigma^2=0.01$.}
\label{table:table20}
\end{table}

\subsubsection{Sequential-mode distributed multi-speaker voice activity detection}
The performance of the proposed detector in the sequential-mode, i.e., SDMVAD+ is evaluated on the challenging multi-speaker scenario with seven active sources, as given in Fig.~\ref{fig:scenario} for K-medoids.

\begin{figure}[H]
\centering
\normalsize
\includegraphics[width=100mm]{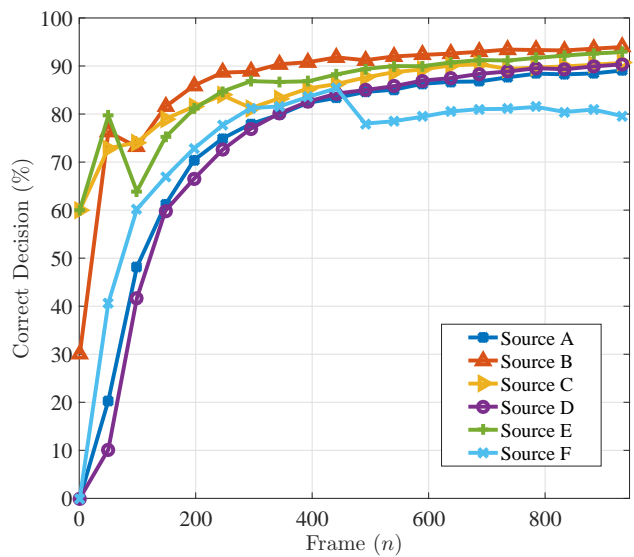}
\caption[Sequential decision for all scenario sources.]{Sequential decision of SDMVAD+ using a growing window.}    
\label{fig:fig17}
\end{figure}

Table \ref{table:table15} displays the VAD results when using a growing window $W^{(n)}, \ n=W^0+1,\ldots,N$ for AWGN of variance $\sigma^2=0.01$. A performance loss of maximally $6 \%$, compared to the batch-mode is ascertained. Figure \ref{fig:fig17} depicts the convergence for the different speech sources to their associated VAD decision. Clearly, the transient behavior of the SDMVAD+ is source-dependent. When using the growing window technique described in Section~\ref{sec:step5}, SDMVAD+ for this setup achieves a steady state performance after approx. $300-500$ speech frames of $30$ms duration each.
\begin{table}[H]
\centering
\normalsize
\begin{tabular}{cccccc}
\toprule
 \multirow{2}{*}{Source} & \multicolumn{5}{c}{Metric}\\
\cmidrule{2-6}
&CD & MD & FA & EER & $C_{\textrm{llr}}^{\min}$\\
\midrule \midrule
 A & 80.2& 5.7 & 14.1 & 0.2 & 0.63 \\
\midrule
 B & 92.9 & 1.7 & 5.4 & 0.07 & 0.32 \\
\midrule
 C & 90.8& \bf{1} & 8.2 & 0.06 & 0.22 \\
\midrule
 D & 90.6 & 7.4 & 2 & 0.07 & 0.31\\
\midrule
 E & \bf{94} & 4.3 & \bf{1.7} & \bf{0.04} & \bf{0.22} \\
\midrule
 F & 89.5& 2.5 & 8 & 0.07 & 0.27 \\
\bottomrule
\end{tabular}
\caption{Proposed sequential-mode VAD (SDMVAD+) with K-medoids using a growing window for signals corrupted by additive white Gaussian noise of variance $\sigma^2=0.01$.}
\label{table:table15}
\end{table}

\begin{figure}[H]
\centering
\includegraphics[width=100mm]{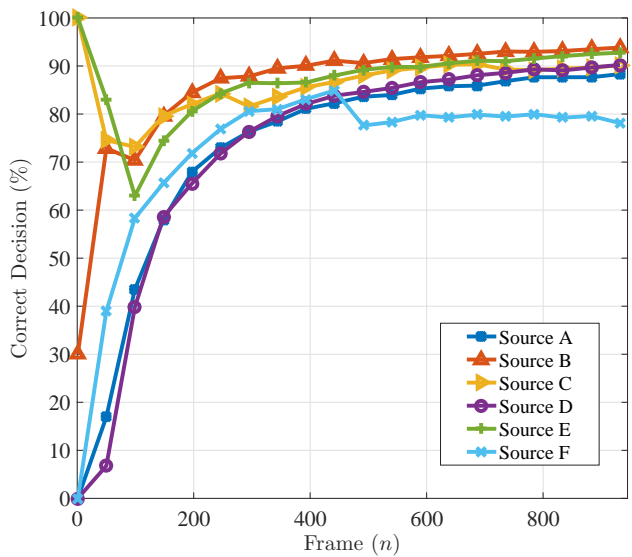}
\caption[Sequential decision for all scenario sources with a fixed sliding window.]{Sequential decision of SDMVAD+ using a fixed sliding window for the scenario given in Fig.~\ref{fig:scenario}.}    
\label{fig:fig18}
\end{figure}

The performance is next analyzed using a fixed size moving window, i.e., the buffer of the past speech data included in the decision is limited to 400 frames. Table \ref{table:table16} summarizes the fixed sliding window SDMVAD+ results, while Fig. \ref{fig:fig18} displays the convergence of the SDMVAD+ for different sources. 
\begin{table}[H]
\centering
\normalsize
\begin{tabular}{cccccc}
\toprule
 \multirow{2}{*}{Source} & \multicolumn{5}{c}{Metric}\\
\cmidrule{2-6}
&CD & MD & FA & EER & $C_{\textrm{llr}}^{\min}$\\
\midrule \midrule
 A & 78.8& 6.1 & 15.1 & 0.2 & 0.65 \\
\midrule
 B & 92.8 & 1.6 & 5.6 & 0.1 & 0.39 \\
\midrule
 C & 90.7& \bf{0.9} & 8.4 & 0.06 & 0.2 \\
\midrule
 D & 90.3 & 7.2 & 2.5 & 0.09 & 0.37\\
\midrule
 E & \bf{93.9} & 4.7 & \bf{1.4} & \bf{0.04} & \bf{0.23} \\
\midrule
 F & 88.9& 2.3 & 8.8 & 0.06 & 0.26 \\
\bottomrule
\end{tabular}
\caption{Proposed sequential-mode VAD (SDMVAD+) with K-medoids using a fixed sliding window for a mixture of energies corrupted by additive white Gaussian noise of variance $\sigma^2=0.01$.}
\label{table:table16}
\end{table}

\section{Conclusions}
\label{sec:conclusions}
A distributed multi-speaker VAD (DM-VAD) method for WASNs has been proposed that does not require a fusion center or prior knowledge about the node positions, microphone array orientations or the number of observed sources. A distributed source-specific energy signal unmixing method, which contains a source-specific node clustering method to locate the nodes around each source (LONAS) as well as a distributed audio source enumeration method (LAPPO) have been introduced. The VAD has been approached by extracting features from the LONAS-MNICA unmixed energy signals by applying K-means type clustering algorithms. All steps of our method showed promising performance compared to existing benchmark methods, wherever possible. More than $85 \%$ of correct decision in the worst case has been obtained for a challenging scenario where 20 nodes observe 7 sources in a simulated reverberant rectangular room. The proposed method is also able to operate for streaming data taking into account a small performance loss 
compared to batch-mode.

\bibliographystyle{plain}
\bibliography{references}

\end{document}